\documentclass[10pt,journal,compsoc]{IEEEtran}
\ifCLASSOPTIONcompsoc
  \usepackage[nocompress]{cite}
\else
  \usepackage{cite}
\fi
\usepackage[dvips]{graphicx}
\usepackage{amsmath}
\usepackage{algorithmic}
\usepackage{algorithm}
\usepackage[caption=false,font=footnotesize]{subfig}
\usepackage{url}
\usepackage{amssymb}
\usepackage{amsthm}
\usepackage{tabularx}
\usepackage{cases}
\usepackage{mathtools}
\usepackage{bbm}
\usepackage{xcolor}
\usepackage{color}
\usepackage{pbox}
\usepackage[binary-units]{siunitx}
\usepackage[utf8]{inputenc}
\usepackage{nicefrac}
\usepackage{amsmath}  
\usepackage{soul}
\newtheorem{remark}{Remark}

\newtheorem{assumption}{Assumption}
\theoremstyle{definition}
\newtheorem{definition}{Definition}
\DeclareMathOperator*{\minimize}{minimize}
\begin{document}
\title{Computation Offloading in Heterogeneous Vehicular Edge Networks: \\On-line and Off-policy Bandit Solutions}

\author{Arash~Bozorgchenani,~\IEEEmembership{Member,~IEEE,} Setareh~Maghsudi,\\ Daniele~Tarchi,~\IEEEmembership{Senior~Member,~IEEE,} and~Ekram~Hossain,~\IEEEmembership{Fellow,~IEEE}%
\IEEEcompsocitemizethanks{ 
\IEEEcompsocthanksitem  A. Bozorgchenani was with the Department of Electrical, Electronic and Information Engineering, University of Bologna, Italy. He is now with the Department of Computing and Communications, Lancaster University, the UK.\protect\\
\IEEEcompsocthanksitem  D. Tarchi is with the Department of Electrical, Electronic and Information Engineering, University of Bologna, Italy.\protect\\
\IEEEcompsocthanksitem S. Maghsudi is with the Department of Computer Science, University of Tübingen, Germany.\protect\\
\IEEEcompsocthanksitem E. Hossain is with the Department of Electrical and Computer Engineering, University of Manitoba, Canada.} %
\thanks{E. Hossain's work was supported by a Discovery Grant from the Natural Sciences and Engineering Research Council of Canada (NSERC).} \thanks{A major part of the work was done during A. Bozorgchenani's visit to the University of Manitoba, Canada, supported by a Marco Polo Scholarship funded by the University of Bologna, Italy, and by the project ``GAUChO - A Green Adaptive Fog Computing and Networking Architecture'' funded by the MIUR Progetti di Ricerca di Rilevante Interesse Nazionale (PRIN) Bando 2015 - grant 2015YPXH4W\_004. The work of S. Maghsudi was supported by Grant 01IS20051 from the German Federal Ministry of Education and Research (BMBF).}
\thanks{*This work has been accepted at IEEE Transactions on Mobile Computing.
Copyright may be transferred without notice, after which this version may no longer be accessible}
}
\IEEEtitleabstractindextext{
\begin{abstract}
With the rapid advancement of Intelligent Transportation Systems (ITS) and vehicular communications, Vehicular Edge Computing (VEC) is emerging as a promising technology to support low-latency ITS applications and services. In this paper, we consider the computation offloading problem from mobile vehicles/users in a heterogeneous VEC scenario, and focus on the network- and base station selection problems, where different networks have different traffic loads. In a fast-varying vehicular environment, computation offloading experience of users is strongly affected by the latency due to the congestion at the edge computing servers co-located with the base stations. However, as a result of the non-stationary property of such an environment and also information shortage, predicting this congestion is an involved task. To address this challenge, we propose an on-line learning algorithm and an off-policy learning algorithm based on multi-armed bandit theory. To dynamically select the least congested network in a piece-wise stationary environment, these algorithms predict the latency that the offloaded tasks experience using the offloading history. In addition, to minimize the task loss due to the mobility of the vehicles, we develop a method for base station selection. Moreover, we propose a relaying mechanism for the selected network, which operates based on the sojourn time of the vehicles. Through intensive numerical analysis, we demonstrate that the proposed learning-based solutions adapt to the traffic changes of the network by selecting the least congested network, thereby reducing the latency of offloaded tasks. Moreover, we demonstrate that the proposed joint base station selection and the relaying mechanism minimize the task loss in a vehicular environment.
\end{abstract}
}
\maketitle
\IEEEraisesectionheading{\section{Introduction}}
\IEEEPARstart{D}{uring} the past few years, edge computing has emerged as a distributed computing paradigm that brings the capabilities and resources of the cloud towards the network edge \cite{SurveyMECMao}. Mobile Edge Computing (MEC) or, as renamed by ETSI, multi-access edge computing, offers an ultra-low latency environment with high bandwidth and real-time access to network resources in a mobile network \cite{MECSurveyTaleb,MEC-Focus}.

Vehicles have been evolving since the second industrial revolution and their role in modern life is imperative. With the rapid technological advancements in Intelligent Transportation Systems (ITS) technologies, vehicles are equipped with wireless communication capabilities for both intra-vehicle and inter-vehicle communications. ITS technologies support a plethora of applications including those for road safety, smart transportation, and location-dependent services \cite{ConnectedVehicles}. Vehicular Edge Computing (VEC) has been widely discussed in the literature~\cite{VANET_Bitam, VaaR2015}, where the computing infrastructures at both the network and the vehicles are used by mobile users. In essence, VEC combines the concepts of vehicular networking and MEC. If performed suitably, task offloading reduces the energy consumption and speeds up the response time of applications in a VEC scenario~\cite{MCCZhang15, OffVANETLyu17, ComOffWang17}.

In vehicular networks, Road Side Units, referred to as Base Stations (BSs) throughout this paper, provide reliable wireless access in their coverage ranges. The edge computing servers for the mobile vehicles are supposed to be co-located with the BSs. We study a task offloading problem from vehicles to the edge computing servers, where the tasks generators can be the driving systems or the passengers of the vehicles. We consider a multi-operator heterogeneous scenario consisting of several access networks with different characteristics and several BSs with different coverage areas. Due to the mobility of the vehicles/users, the traffic load across the networks is time-varying. In such a dynamic environment, vehicles select the best network to perform computation offloading. This might yield intensive traffic load and congestion at the edge server-side, thereby harming the offloading performance. This paper proposes an offloading decision framework that can dynamically select the most suitable network.

Task offloading in VEC involves several challenges. For example, a low latency, which is crucial to ensure an acceptable Quality of Experience (QoE), is difficult to achieve. In a computation offloading scenario, the latency consists of different components including the communication latency, the processing latency, and the waiting time at the computing buffers. In MEC, the waiting times at the edge computing servers often dominate the transmission times~\cite{WaitingTimeBoxma}; therefore, we concentrate on the analysis of waiting time for an offloaded task, referred to as the {\em latency of computation offloading}.  

 For a reliable offloading decision, the vehicle should be aware of both task and network parameters. While task size and number of operations for processing can be apriori determined by the vehicle, the waiting time at the edge server-side, depending on the network load, is unknown. This lack of knowledge weakens the decision on task offloading. 

The VEC scenario is characterized by frequent changes in network availability and traffic, mainly due to vehicle mobility and the random characteristics of the wireless medium. In this work, we focus on a non-stationary, or more specifically, a {\em piece-wise stationary} scenario, where the traffic load of the networks remains constant within some period, and changes at some unknown time instant called {\em change points}. The {\em piece-wise stationary} scenario is a suitable model to formalize the non-stationarity of the vehicular environment in terms of network congestion and data traffic. In such a scenario, detecting change points, which potentially alter the offloading decisions, is challenging. To develop efficient decision-making policies, the availability of a data set, including the previous offloading decisions, the network's characteristics, and the traffic's statistics for each network, can be quite useful. Machine learning is a key method for extracting and learning useful information from data to develop decision-making policies \cite{XingBigData15}. Motivated by the above-mentioned factors, i.e., the time-variant statistical characteristics of the vehicular environment, lack of perfect knowledge before decision-making, and the inevitability of exploration-exploitation trade-off in such a dynamic environment, we propose two machine learning-based methods for making decision on offloading. Among several machine learning methods, we focus on \textit{multi-armed bandit (MAB)} theory, as it enables the user/vehicle to learn through time by exploiting the bandit feedback. In bandit theory, the decision-maker observes some context, performs an action, and receives some feedback in terms of cost or utility. The decision-maker then uses the history of action-outcome pairs to improve its future performance. 

The policies used in such an interactive environment can be deployed either on-line or off-line manners, each having its pros and cons \cite{OffPolicyJagerman19}. In particular, the on-line approach is suitable for scenarios where no prior information is available to the decision-maker. Such an approach, however, might require excessive time for convergence to the optimal decision. To implement the off-policy method, the existence of the data set is necessary. Such a data set shall improve the performance by enhancing the decision-making policy. The improvement depends on several factors such as the quality of the data, the performance of the underlying decision-making policy, the problem setting such as the dynamics of the environment, and the like.

 In this paper, we propose an on-line and and an off-policy algorithm each with two phases: decision making on where to offload, and task offloading. In both the algorithms, the vehicle (or the vehicular user) takes advantage of the historical offloading records. While in the on-line algorithm, the vehicle adapts its decisions over time, in the off-policy approach, the decision is made once for a given amount of time. Using the proposed solutions, a vehicle is able to perform an efficient network selection for computation offloading in presence of unpredictable or non-stationary traffic load at the edge computing servers. The main contributions of this paper can be summarized as follows:
\begin{itemize}
 \item We model the computation offloading problem in a VEC environment by considering a {\em piece-wise stationary scenario}, which is a good approximation of the dynamicity in a vehicular network. To the best of our knowledge, no previous work in the literature has investigated the task offloading problem in a piece-wise stationary VEC scenario. 
\item  We develop an on-line network selection scheme based on congestion and traffic patterns in a multi-access edge computing network using the MAB theory, which is a suitable mathematical framework for problems with no prior information and limited feedback. The proposed solution aims at minimizing the latency for the offloaded tasks.
\item We propose an off-policy network selection approach by exploiting the historical offloading data set. The proposed approach first detects the change points of the piece-wise stationary scenario. Then, the network selection is performed based on the developed off-policy approach aiming at minimizing the latency for the offloaded tasks. To the best of our knowledge, this is the first work to study off-policy learning for task offloading in a VEC environment.
\item  We propose a BS selection method based on the sojourn time of the vehicle in the selected network. In addition, we propose a relaying mechanism to minimize the probability of task loss during the offloading procedure due to mobility of the vehicles. The relaying mechanism regards the task size and the application types as random variables that provides generality to the solution.
\item We perform extensive numerical analysis to demonstrate the effectiveness of the proposed approaches to adapt to the changes in traffic load in the considered piece-wise stationary scenario. Adapting to the environment reduces latency while guaranteeing a tolerable task loss. 
\end{itemize}
The rest of the paper is organized as follows. In Section~\ref{sec:ReWorks}, we review the state of the art. In Section~\ref{sec:model}, we describe the system model. Section~\ref{Net_Selection} introduces our proposed on-line and off-policy network selection approaches. In Section~\ref{BS_Relay}, we introduce the BS selection and relaying methods. In Section~\ref{sec:numericalresult}, we present the performance evaluation results. Section~\ref{conclusion}, concludes the paper. 
\section{Related Works}
\label{sec:ReWorks}
Computation offloading in a mobile environment has been investigated intensively in recent literature. In \cite{MCCZhang15}, the authors study the problem of energy conservation on mobile devices by offloading tasks to the infrastructure-based cloud. Reference \cite{OffVANETLyu17} proposes a heuristic offloading decision algorithm that optimizes the offloading decision while allocating communication and computation resources. A similar work is  \cite{ComOffMixedFogCloud}, where the authors address the joint optimization of the offloading decisions, the allocation of computation resources, transmit power, and radio bandwidth. The problem of joint optimization of offloading decision, resource allocation, and content caching strategy is considered in \cite{ComOffWang17}. In \cite{LTEMixedFogCloud}, the authors target a joint optimization of offloading decision making, computation resource allocation, resource block assignment, and power distribution in a mixed fog and cloud network. In \cite{ArashWCNC}, the authors propose a partial offloading solution considering the sojourn time of the mobile devices in order to minimize the latency and task loss. Reference \cite{NetSelINFOCOM19} couples network selection with service placement. The authors propose an online heuristic, where a centralized operator at each time instant chooses a base station and a proper service placement for each user. In \cite{YuJSAC16}, the authors consider a predictive Lyapunov optimization technique and propose a greedy predictive energy-aware network selection and resource allocation scheme. They model a mobile environment where the operator can adjust between power consumption and traffic delay.

In~\cite{AdaptiveComOff16}, the authors consider a dynamic environment with different wireless networks among which the decision is to select the network reducing the execution cost. Reference \cite{MobMigOff} proposes a fog offloading scheme in a mobile environment. In brief, if the sojourn time of a mobile user is less than the transmission time, it performs a local computation. In case of offloading, if the computation time is less than the sojourn time, the base station sends the result back to the user, otherwise, the task will be migrated to the cloud for relaying the result to the destination BS of the user. The objective of the work is to minimize cloud migration by proposing a generic-based solution. However, no queuing model is considered in these works and the task waiting time for processing is ignored. 

Furthermore, there exists a rich body of the research work that studies the task offloading problem in Vehicular Adhoc Networks (VANETs) and VEC. In \cite{MECVehicular17}, the authors propose a MEC-based computation offloading framework to minimize the vehicles’ cost for task offloading while guaranteeing processing delay. In \cite{VECNash17}, to maximize the economical profit of service providers while maintaining low delays, the authors develop a game-theoretical approach that jointly optimizes the task offloading decision and computation resource allocation. Reference \cite{VFogInfrastracure} proposes to utilize vehicles as the infrastructure for communication and computation. The authors then analyze both scenarios of moving and parked cars as infrastructure. In \cite{VehicuResAll}, the authors formulate a dual-side cost minimization, which jointly optimizes the offloading decision and local CPU frequency on the vehicles' side and the radio resource allocation on the servers' side. A graph-based scheduling scheme for V2V and V2I communication has also been studied in \cite{VehNetScheduling}. Most of the research works described before neglect the waiting time and the reception time when developing models and solutions for network- or BS selection. Moreover, often they ignore the fact that the selected BS might not be able to process the offloaded task within the sojourn time of the vehicle. 

Recently, learning theory has been applied by several authors to address the computation offloading problem. In~\cite{TLRL}, the authors target the minimization of the delay and the utilization of physical machines for task offloading in mobile cloud computing. They propose a two-layer Reinforcement Learning (RL) structure, where a deep RL method is used to select the optimal cluster in the first layer. In the second layer, a Q-learning approach is used to select the optimal physical machine in the selected cluster. However, the offloading is to the cloud and there is no mobility considered in the scenario. Similarly, in \cite{RLFogCom}, the authors aim at minimizing the latency for task assignment to the servers and propose an RL method. However, this work lacks a detailed formulation of the delay model and considers a simplified scenario. In~\cite{MashhadiComNet20}, the authors have formulated user to server allocation as an auction problem and proposed two allocation and payment deep neural networks to maximize the profit of edge servers and satisfy the energy and delay constraints of users.

Reference \cite{LearningOffEdge} utilizes the multi-arm bandit (MAB) theory to minimize delay in computation offloading. The proposed MAB solution is for a Vehicle-to-Vehicle (V2V) scenario and assumes that all offloaded tasks are received from the same vehicle when offloaded; however, due to the (fast) mobility, the vehicles cannot receive the result while they are within the sojourn time of the processing vehicle. Moreover, the system model does not include any queuing model, although every processing vehicle might receive multiple tasks. Besides, the environment is stationary, which is not realistic in a fast-varying vehicular environment. A similar work is \cite{Adaptive_EE}, where the authors consider network load as a parameter in the Upper Confidence Bound (UCB) function such that offloading to new nodes is favorable when the network load is low. By this, when the network load level is low, the exploration is emphasized. The exploitation has a higher weight in heavy load scenarios. Reference \cite{ZhangOnline20} studies the task assignment problem in small cell networks. The authors calculate the probability of assigning the tasks to each network based on bandit feedback. They also propose a greedy collaborative task offloading scheme among small cell networks considering the selection probability. In \cite{BanditMECSun20}, the authors formulate an online optimization problem for task offloading in MEC. They target the queue stability and delay minimization by controlling the portion of data to be analyzed locally and to be offloaded. However, in both of the aforementioned papers, the methods and also the considered scenarios are different from ours.

References \cite{Ghoorchian19:MAB}, \cite{Maghsudi_PieceWise}, and \cite{Nikfar15:MBC} apply piece-wise stationary bandit models, respectively, for server selection, small cell planning, and channel selection for power line communication. The problems under investigation, are, however, not related to the one considered in this paper. Task allocation under uncertainty is also considered in \cite{Maghsudi19:DTM}; however, the approach is one-shot based on the expected utility, rather than being iterative. 

\textcolor{black}{To summarize, several state-of-the-art research assume the availability of knowledge about offloading and network parameters, although such an assumption is not realistic in fast-varying heterogeneous environments. Moreover, most of the existing learning-based offloading solutions do not have suitable policies for decision making in a dynamic vehicular environment.} A large body of the current literature concerning distributed computation offloading in vehicular scenarios ignore the non-stationarity of the medium. That includes, for example, the dynamic variation of traffic load at the edge servers (e.g. due to vehicles' mobility). In this work, we take these non-stationarity and traffic dynamicity into account to develop two on-line and off-policy learning-based solutions. 
\section{System Model}
\label{sec:model}
In this section, we first describe the network model and the offloading environment. Afterwards, we will explain the task computation, communication, and queuing models.
\subsection{VEC Model}
\label{VecModel}
We consider a two-layer architecture for VEC, where there are several Edge Units (EUs) belonging to the set of EUs $\mathcal{U}$, in the first layer and $u_i$ is the EU of interest. All the EUs have certain computational and storage capabilities. The second layer includes $M$ wireless networks, each having certain number of BSs~\cite{AdaptiveComOff16}. $\Im_{m}=\{N_1,\dots, N_M\}$ shows the set of all networks, where $N_m$ represents the $m$-th network. Moreover, $\Im_{j}^m=\{ N_{m_1},\dots,N_{m_J}\}$, shows the set of all BSs of the $m$-th network type, where $N_{m_j}$ identifies the $j$th BS of the $m$th network. Every EU can communicate with a BS by means of cellular-based communications. Each BS is equipped with a number of computational servers and therefore can process different tasks. In general, the BSs have higher computational capabilities compared to the EUs. They can aggregate the EUs' traffic requests and process the offloaded tasks. The BSs of each network {\em type} are homogeneous, meaning that they have the same computational capability, edge server traffic model, and coverage. In contrast, the BS characteristics for different network types vary from one network to another. Note that in a certain coverage area, BSs of different network types (e.g., owned by the same network operator) and also BSs of the same type (e.g., owned by different network operators) exist.

\begin{figure}
\begin{center}
\includegraphics[width=\columnwidth]{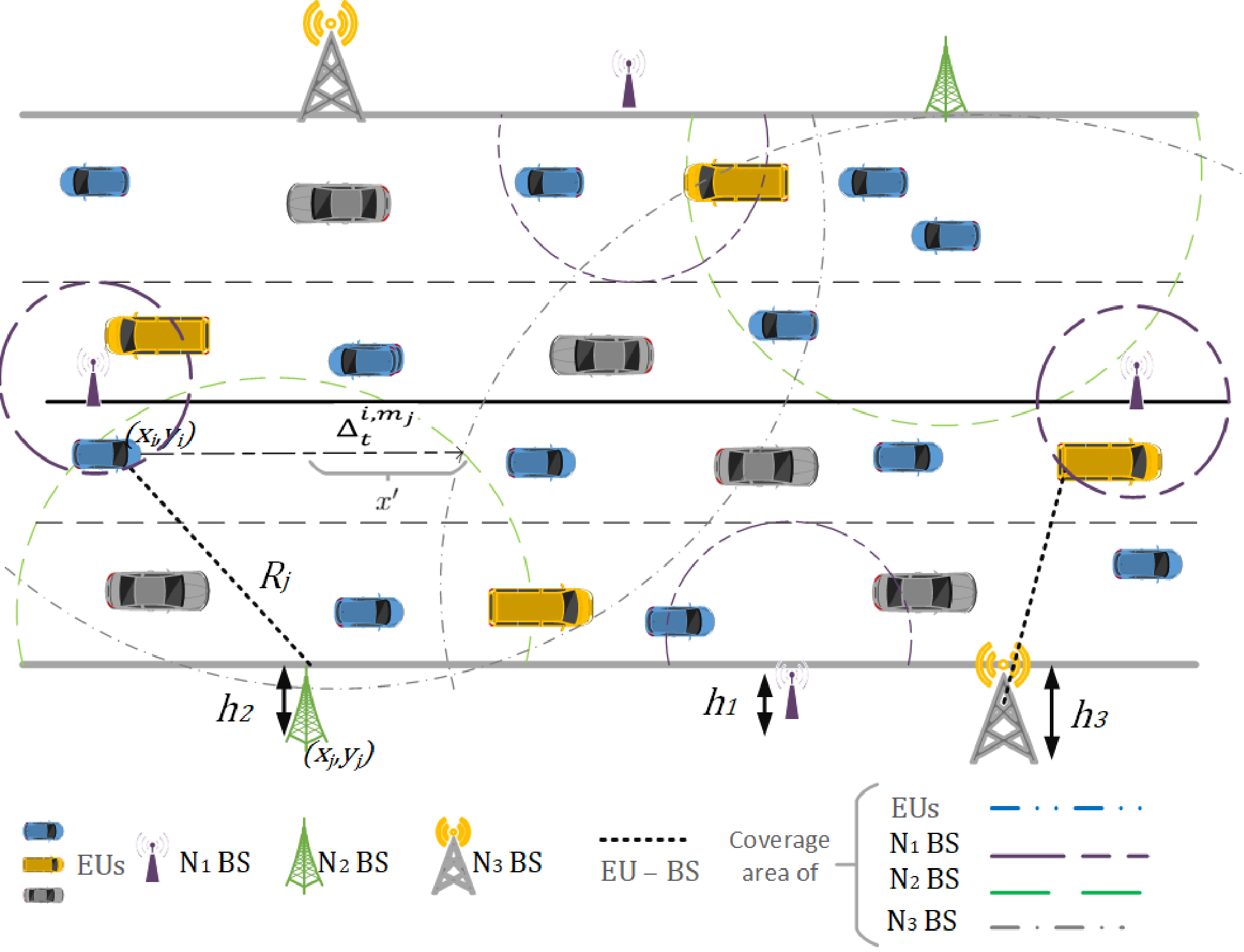}
\caption{Computation offloading in a vehicular edge computing scenario.}
\label{Off_Setting}
\end{center}
\end{figure}

We consider an urban scenario as shown in Fig.~\ref{Off_Setting}, where the vehicles act as EUs. Each vehicle $i$ moves with some velocity $v_i$. The vehicles move either from left to right or in the opposite direction, depending on their location. The speed and the direction of each vehicle remain fixed through time~\cite{DL_VANET18, MECVehicular17}.  
A \textit{task loss} occurs if an offloading vehicle cannot receive the result of the offloaded task due to its mobility, i.e., due to the short sojourn time. In a heterogeneous wireless access network, the BSs of different networks overlap while covering the entire area. In addition to different coverage and computational capacities, heterogeneous networks can also face different traffic demands due to the offloaded tasks and therefore different levels of congestion at the edge servers. We also assume a full task offloading scenario. 

Based on the discussion above, the offloading decision in dynamic vehicular environments is twofold. First, an EU selects the optimal network to offload its task, based on the level of congestion. Afterwards, it selects a BS in the selected network type having a high probability on the reception of the result of the offloaded task. Since in VEC scenarios, latency is mainly dominated by waiting time, in the rest of the paper we focus on waiting time minimization. \textit{Thus, the goal is to address waiting time minimization problem, by a suitable selection of the network and BS for computation offloading.} 
\subsection{Task Computation and Communication Model}
The computation time for the $l$th task generated by the EU is defined as:
\begin{equation}
	\label{eq:TaskGen}
	T^{\text{com}}_l=\begin{cases}
   \frac{O\cdot \rho^{\text{up}}_l}{\eta_i} & \quad \text{local computation} \\
   \frac{O\cdot \rho^{\text{up}}_l}{\eta_{m_j}} & \quad \text{otherwise,}
	\end{cases}
\end{equation}
where $O$ represents the number of operations required for computing one byte, $\rho^{\text{up}}_l$ is the task size of the $l$th task generated by the EU in byte, and $\eta_*$ is the Floating-point Operation Per Second (FLOPS) depending on the CPU of the device, which could be either a local processor or a processor at the $m_j$th BS. 

The transmission time of the $l$th task from the EU to the $m_j$th BS is given by
\begin{equation}
T^{\text{trans},{m_j}}_{l,t}=\frac{\rho^{\text{up}}_l}{r_{m_j,t}^{\text{up}}},
\end{equation}
where $r_{m_j,t}^{\text{up}}$ is the up-link data rate of the link between the EU and the $m_j$th BS at time instant $t$. The result of the processed task should be sent back from the $m_j$th BS to the EU, leading to a reception time as
\begin{equation}
T^{\text{rec},{m_j}}_{l,t}=\frac{\rho^{\text{dl}}_l}{r_{m_j,t}^{\text{dl}}},
\end{equation}
where $r_{m_j,t}^{\text{dl}}$ is the down-link data rate and $\rho^{\text{dl}}_l$ is the size of the result of processing. 

We assume that there is no inter-cell interference in the up-link transmissions (e.g. due to orthogonal channel sets used by different BSs), whereas the EUs may experience intra-cell interference. We make the following assumption.
\begin{assumption}
The EU $u_i$ inside a cellular network experiences independent and identically distributed interference signals caused by other EUs. However, inter-cell interference can be neglected due to inter-cell coordination. 
\end{assumption}
In a dense vehicular network, a fully-orthogonal transmission is often not possible due to mobility and density; consequently, we consider intra-cell interference (as in \cite{IntraCellInterference11}). It is worth noting that our approach is generic and applicable also in combination with other interference models. Focusing on a specific time instant when EU $i$ is inside the $m_j$th cell, we define $\mathbb{I}^{i,\iota}=1$, where $i, \iota \in \mathcal{U}, i \neq \iota$, if the $i$th EU and $\iota$th EU cause interference to each other. Thus the interference received by the $i$th EU yields
\begin{equation}
\label{Interference}
I^i_{m_j,t}= \sum_{\iota} \mathbb{I}^{i,\iota}\beta^{\iota,m_j}_t P^{\iota}_{\text{trans}},
\end{equation}
where $\beta^{\iota,m_j}_t$ is the path-loss attenuation factor. It is a function of the distance between the BS and the vehicle, shown as $d^{i,{m_j}}_t$. Moreover, $P^{\iota}_{\text{trans}}$ is the transmission power of the interferer. The distance between the $i$th EU and the $m_j$th BS is given by 
\begin{equation}
\label{distance}
d^{i,{m_j}}_t=
\sqrt{h_{m_j}^2+|y_{m_j}-y_{i,t}|^2+|x_{m_j}-x_{i,t}|^2},
\end{equation}
where $\{x_{i,t},y_{i,t}\}$ and $\{x_{m_j},y_{m_j}\}$ are the positions of the $i$th EU at time $t$ and the $m_j$th BS, respectively. Moreover, $h_{m_j}$ is the height of the $m_j$th BS.

If $B_m$ represents the bandwidth of a channel, the transmission rate in the up-link from the $i$th EU to the $m_j$th BS at time instant $t$ can be given as
\begin{equation}
\label{upRate}
r_{m_j,t}^{\text{up}}=B_m\log_2\left(1+\frac{\beta^{\iota,m_j}_t P^i_{\text{trans}}}{I^i_{m_j,t}+\zeta}\right),
\end{equation}
where $P^i_{\text{trans}}$ is the transmission power of the $i$th EU, $\zeta$ is the noise power defined as $\zeta=N_0 B_m$, where $N_0$ is the noise power spectral density. When sending the processed tasks back to the vehicles, the transmission rate in the down-link, i.e., from the $m_j$th BS to the $i$th EU, is calculated similarly  using the BS transmission power except that there will be no interference due to the orthogonal channels used by different BSs. We assume that the channel is quasi-static during the transmission- and reception periods.
\subsection{Traffic Model at Edge Computing Servers}
We consider a Poisson-Exponential (PE) queuing model~\cite{Pois_Exp}, as described in the following. 
 Servers have a buffer large enough to queue all tasks. The arrival of tasks to the BSs that belong to the $m$th network follows a Poisson distribution $\operatorname{Pois}(\lambda_{m})$. Moreover, the service time for each task follows an exponential distribution $\operatorname{Exp}(\mu_{m})$. Let $\vartheta_t^m(\lambda_{m},\mu_m)$ denote the amount of tasks in the queue of the $m$th network at time instant $t$. {$T_{w_{m}}^l(\vartheta_t)$}\footnote{More specifically, a task is offloaded to a BS, $m_j$, in the network $m$; however, for the sake of simplicity of the notation we omit the BS index.} shows the total waiting time of the $l$th task offloaded to $m$-th network.
%
%
The BSs belonging to the same network type are assumed to be statistically identical, meaning that they have the same task-arrival and departure parameters. 

Computational offloading involves two phases: (i) Network selection based on congestion characteristics; and (ii) BS selection in the selected network based on the sojourn time. In the following, we describe our proposed network and BS selection procedures.
\section{Network Selection}
\label{Net_Selection}
Since we assume that the BSs of each network have the same statistical parameters, they suffer the same congestion level on average, and therefore, the first step of the offloading decision boils down to network selection. In this section, we formulate the network selection problem with the aim of minimizing the task waiting time at the edge servers. We then propose two learning approaches. In the first approach, an on-line UCB algorithm is developed, while in the second approach, an off-policy method is developed to select the best network by incorporating a change point detection mechanism. For a better clarity, in the following, we list the key notations of this section in Table~\ref{notations}. 

\begin{table}[tbp]
	\centering
	\caption{Nomenclature (in the order of appearance)}
	\label{notations}
	\begin{tabular}{ | m{0.14\columnwidth} | m{0.76\columnwidth}| } 
		\hline
		\textbf{Notation} & \textbf{Description} \\
		\hline
		$\mathbb{Q}_t^m$ &  Indicator function for offloading to network $m$ \\
		\hline
		 $T_{w_{m}}^l$ & Waiting time for task $l$ in network $m$   \\
		\hline
		$c^m_t$	& Instantaneous cost function \\
		\hline
		 $\lambda_m$ & Mean arrival rate of the tasks in network $m$ \\ 
		\hline
		 $\mu_m$ & Mean service time of the tasks in network $m$ \\ 
		\hline
		 $\tau$ & UCB window length \\
		\hline
		 $C^m_t(\tau)$ &  Number of selection of network $m$ in $\tau$ \\
		\hline
		$C_t$ & Number of offloaded tasks by round $t$ \\
		\hline
		 $\hat{c}^m_t(\tau)$ & SW-UCB cost index \\
		\hline
		$\bar{c}^m_t(\tau)$ &  Average accumulated cost\\
		\hline
		$\xi$ &  A constant weight on SW-UCB index\\
		\hline
		$\beta$ & Upper bound on exploration factor on SW-UCB index\\
		\hline
		$L_t^*(m)$ &  Expected cost of offloading to the optimal network\\
		\hline
		$L_t(m)$ & Expected cost of offloading to the selected network \\
		\hline
		$R_T^{\text{On}}$ & Cumulative regret of on-line approach \\
		\hline
		 $\mathcal{B}$ & Set of change points\\
		\hline
		 $b_\upsilon$ &  $\upsilon$-th change point\\
		\hline
		$\mathcal{Z}$ & Set of intervals \\
		\hline
		$z_{\psi}$ & $\psi$-th interval\\
		\hline
		$\omega$ & Length of an interval\\
		\hline
		$\nu^{\psi}$ & Vector context for each interval in the logs \\
		\hline
		$\chi$ & Context distribution\\
		\hline
		$D^{\psi}$ & Cost distribution in each interval\\
		\hline
		$\tilde{c}_t$ & Expected cost of selected action at a time\\
		\hline
		$\mathcal{H}^{\psi}$ & Logs in $\psi$-th interval\\
		\hline
		 $\mu_b$ & Inter-arrival of change points \\
		 \hline
		$\delta$ & Error range in the inter-arrival of change points \\
		\hline
		$\bar{t}(b_{\upsilon})$	& Mean occurrence time of a change point  \\
		\hline
		$H_0$ & Null hypothesis \\
		\hline
		$H_1$ & Alternative hypothesis  \\
		\hline
		 $\varsigma$ &  Change point moment\\
		\hline
		$\lambda(X)$ & Likelihood ratio of the samples\\
		\hline
		$\mathcal{L}(D|X_i)$ & Likelihood function\\
		\hline
		$\hat{\mu}$ & Mean of samples in $H_0$\\
		\hline
		$\hat{\mu}_0$ & Mean of samples before change point in $H_1$\\
		\hline
		$\hat{\mu}_1$ & Mean of samples after change point in $H_1$\\
		\hline
		$\sigma^2$ & MLE for variance\\
		\hline
		$\alpha$ & Significance level\\
		\hline
		$\mathcal{D}$ & Historical data in all intervals\\
		\hline
		$\pi_0^{\psi}$ & Logging policy in $\psi$-th interval\\
		\hline
		$\pi_w^{\psi}$ & Target policy in $\psi$-th interval\\
		\hline
		$V^*(\pi_w^\psi)$ & Optimal value of the target policy in interval $\psi$\\
		\hline
		$\hat{V}_{\text{IPS}}(\pi_w^\psi)$ & Estimated value of the target policy with IPS estimator\\
		\hline
		$R_T^{\text{Off}}$ & Cumulative regret of off-policy approach\\
		\hline
	\end{tabular}
	
\end{table}

\subsection{Problem Formulation for Network Selection}
%
%
%
 We consider a time-slotted system, where time slot $t$ starts at time instant $t$. We assume that tasks are always generated at the beginning of a time slot. Then each time slot becomes a decision round. Hence, we denote both time slot and decision round as $t$. Let $\mathbb{Q}_t^m$ be an indicator function that returns 1 if EU offloads to the $m$th network at the decision round of $t$. During $T$ rounds of offloading decision, the EU's objective is to offload a task $l$ to a network such that its expected waiting time is minimized. Therefore, the problem can be stated as follows:
\begin{equation}
\label{Obj-Goal}
\mathbf{P1}:
\displaystyle{\minimize_{m \in \Im_{m}} \Bigg\{ \sum_{t=1}^{T}   T_{w_{m}}^l(\vartheta_t^m) \Bigg\}}
\end{equation}
subject to
\begin{align}
&\mathbf{C1}: \sum_{m \in \Im_{m}} \mathbb{Q}_t^m = 1, \quad \forall t \label{c1}
\end{align}
Constraint \eqref{c1} guarantees that each EU offloads to only one network. The actions of other users/agents are abstracted in the queuing process; hence, $\vartheta_t^m$ models the traffic generated by other EUs in each network.

The waiting time for an offloaded task depends on the load of the selected network. In case of the availability of network load information, the EU selects the network as $\operatorname*{argmin}_{m} \{ T_{w_{m}}^l(\vartheta_t^m)\}$. However, the EU is not aware of the offloading decisions of other EUs and therefore the BSs' queue status. Therefore, in the following, we develop a learning-based solution for network selection for computation offloading in a VEC environment.
\subsection{An On-line Learning Solution for Network Selection} 
The latency for a task depends on several parameters related to the EU and the BSs. An EU knows the task parameters such as the size and the required number of process operations. However, the traffic load in each network is unknown to the EU as it depends on the vehicles' arrivals and departures and the offloading demands in the corresponding coverage area. To this aim, in our latency minimization problem, we focus on task waiting time that depends on traffic load in each network. We utilize the single-player MAB model, which is suitable to solve the problems with limited information such as $\mathbf{P1}$. 

In a bandit model, an agent gambles on a machine with a finite set of arms. Upon pulling an arm, the agent receives some instantaneous reward from the reward generating process of the arm, which is not known a priori. Since the agent does not have sufficient knowledge, at each trial it might pull some inferior arm in terms of reward which results in some instantaneous regret. By pulling arms sequentially at different trials of the game, the agent aims at satisfying some optimality conditions \cite{Maghsudi_PieceWise}. Since our objective is to minimize the waiting time, we opt to use the notion of \textit{cost} instead of reward. Therefore, the goal is to minimize the cost. In brief, in our model:
\begin{itemize}
\item The EU and the networks represent the agent and the arms, respectively.
\item The instantaneous loss of pulling arm is the difference between the expected waiting time and the waiting time corresponding to the optimal arm.
\end{itemize}
At every round, the player selects an arm (i.e., a network), for offloading a task, observes its loss, and updates the estimation of its loss distribution. Each time a network is selected, the player observes the waiting time that is used for cost calculation. The objective is to minimize this loss brought by wrong network selection over time. We define the instantaneous cost function for choosing action $m$ (network selection) at round $t$ as
\begin{equation}
\label{Reward}
c^m_t=\Bigg\{
 T_{w_{m}}^l(\vartheta_t) \Bigg\}\cdot \mathbbm{1}_{ \{\mathbb{Q}_t^m=1\}}\\
\end{equation}
The offloading latency depends largely on the task queuing time; however, due to the dynamicity of a vehicular network such as vehicles' density, often no information is available about this variable. Moreover, the statistical characteristics of vehicles' arrival and density, as well as offloaded tasks generation distribution, change over time. Hence we make the following assumption:
\begin{assumption}
\label{Piece-wise}
For all BSs in network $m$, $\lambda_m$ and $\mu_m$ do not remain constant; rather, they remain fixed over specific periods of time and change from one period to another.
\end{assumption}
Due to the aforementioned assumption, the queue status of the BSs is piece-wise stationary, where the length of the period and the distribution are unknown. As mentioned in Section~\ref{VecModel}, for the BSs that belong to the same network type, the average rates of arrival and departure of the tasks are assumed to be the same, thereby have similar {\em average} queuing behavior. This assumption simplifies the learning process\footnote{Note that, although the average rates are the same, the instantaneous values can differ.}.

Network selection for task offloading with MAB is a sequential optimization problem. The previously offloaded tasks provide latency/cost information. However, this information may not be accurate due to insufficient sampling of each arm. Hence, an exploration-exploitation trade-off shall be addressed. 
One of the most seminal policies to address the exploration-exploitation trade-off is Upper the Confidence Bound (UCB) algorithm \cite{UCB1}. In the UCB algorithm, at every round of the game, an index is calculated for each arm corresponding to the average reward of pulling the arm in all previous rounds (the exploitation factor) and the tendency in pulling the arm for another round (the exploration factor). The UCB policy considers the entire reward history to calculate the arms' indexes; however, in a piece-wise stationary setting, the old observations are less important \cite{hartland_MetaBandit}. Hence, to calculate the arms' indexes, it would be beneficial to disregard the obsolete observations and consider only the last $\tau$ observations. In our vehicular scenario, we exploit \textit{Sliding-Window UCB} (SW-UCB) \cite{D-UCB&SW-UCB} algorithm that uses the last $\tau$ observations for learning, as described in the following.

The number of times the $m$th arm has been selected during a window with length $\tau$ up to round $t$ is given by
\begin{equation}
\label{N_Off_arm}
C^m_t(\tau)=\sum_{s=t-\tau+1}^{t} \mathbbm{1}_{ \{\mathbb{Q}_s^m=1\}}.
\end{equation}
Let us define the total number of offloaded tasks by the EU, $C_t$, by round $t$ to all the selected arms (networks) as
\begin{equation}
\label{N_Off_Task}
  C_t=\sum_{m \in \Im_{m}}\sum_{s=1}^{t} \mathbbm{1}_{ \{\mathbb{Q}_s^m=1\}}.
\end{equation}
Inspired by the SW-UCB, we define the cost index of pulling arm $m$ at round $t$ as
\begin{equation}
\label{UCB}
\hat{c}^m_t(\tau)=\bar{c}^m_t(\tau)-\beta \sqrt{\frac{\xi\log(\min\{C_t,\tau\})}{C^m_t(\tau)}}.
\end{equation}
At every decision-making round, the agent pulls the arm with the minimum $ \hat{c}^m_t(\tau)$. In (\ref{UCB}), the first term on the right side corresponds to the exploitation factor, since $\bar{c}^m_t(\tau)$ is the average accumulated cost up to round $t$ with window length $\tau$. Formally,

\begin{equation}
\bar{c}^m_t(\tau)=\frac{1}{C^m_t(\tau)}\sum_{s=t-\tau+1}^{t}  c^m_s.     
\end{equation}
The second term on the right side of (\ref{UCB}) is the exploration factor, where $\xi$ 
is a constant weight and $\beta$ is an upper bound on exploration factor. 

Let $L_t^*(m)=\min_m \mathbb{E} \left[ c^m_{t} \right]$ 
represent the expected cost of offloading to the optimal network, while $L_t(m)=\mathbb{E} \left[\hat{c}^m_t(\tau)\right]$ denotes the expected cost of offloading to the $m$th network selected by the proposed MAB method. We define the cumulative regret during $T$ rounds as
\begin{equation}
\label{Regret_on}
R_T^{\text{On}}= \sum_{t=1}^{T} L_t(m) - \sum_{t=1}^{T} L^*_t(m),    
\end{equation}
which is the expected loss of the algorithm compared with the optimal network selection.
\subsection{An Off-Policy Learning Solution for Network Selection}
In this section, we address the problem of network selection by proposing an off-policy learning method in a bandit setting. In off-policy learning, the goal is to estimate the value of a target policy  exploiting a historical (or logging) policy. Off-policy learning can be seen as a parameterized policy such as weights in a neural network. In this setting, each request from the EU provides a context, based on which the system selects an action and incurs some cost. Such contextual-bandit data can be logged in large quantities and used for future purposes as training data~\cite{Unbiased_Learning16}.
Different from the on-line learning setting, off-policy learning is statistically more challenging
since the collected logs are generated by a logging policy that differs from the current policy to be developed \cite{SurrogatePolicy18}. 

A large body of literature considers a stationary environment for off-policy evaluation; in this work, however, we consider a non-stationary environment. That is, in some time intervals both cost distribution and context remain stationary, though the distribution might change at some unknown time. Therefore, in this context, the problem of learning becomes two-fold: (i) estimating the occurrence time of the change points, and (ii) developing a target policy for each stationary interval based on the post-change distribution.
\subsubsection{Change Point Detection}
\label{ChangepointSection}
In an on-line setting, it is vital to detect a change as fast as possible while minimizing the rate of false alarms. In an off-line scenario, the goal is to identify the patterns or distribution of the change point occurrence based on the observed logs. For the formulated VEC problem, we propose a mixed off-line and on-line strategy for change point detection. 

Let  $\mathcal{B}=\{b_1,\ldots,b_\upsilon,\ldots,b_\Upsilon \}$ be the set of change points, and $\mathcal{Z}=\{z_1,\ldots,z_\psi,\ldots,z_\Psi \}$ the set of intervals, where $\Psi=\Upsilon+1$. Each interval $z_{\psi}$ has an unknown duration including $\omega$ rounds.

The input data consists of a finite streams of tuple  $\left\langle\nu^{\psi}, \tilde{c}_t\right\rangle$ where $\nu^{\psi}$ is a context vector, $\nu^{\psi}=[\lambda_m^{\psi}, \mu_m^{\psi}]_{m=1}^M$, and $\lambda_m^{\psi}$ and $\mu_m^{\psi}$ are the $m$th network task generation and service rate distribution parameters in the interval $\psi$. The context $\nu^\psi$ is a feature vector drawn according to some unknown distribution $\chi$ during each interval. If we assume that $c^m_{t}$ in \eqref{Reward} can be characterized by an interval-dependent {distribution}\footnote{It should be noted that, in a non-stationary environment, both cost and context distribution change after each change points.} conditioned on $\nu^\psi$ and $m$, denoted by $D^{\psi}(c_t^m|\nu^\psi,m)$, we can define $\tilde{c}_t=\mathbb{E}\left[ c^m_{t} \right]$
as the expected cost of the selected action at every time instant. Since an action implies the selection of a certain network, in the expectation we remove the dependency on the selected network.

\begin{figure}[h]
\begin{center}
  \includegraphics[width=\columnwidth]{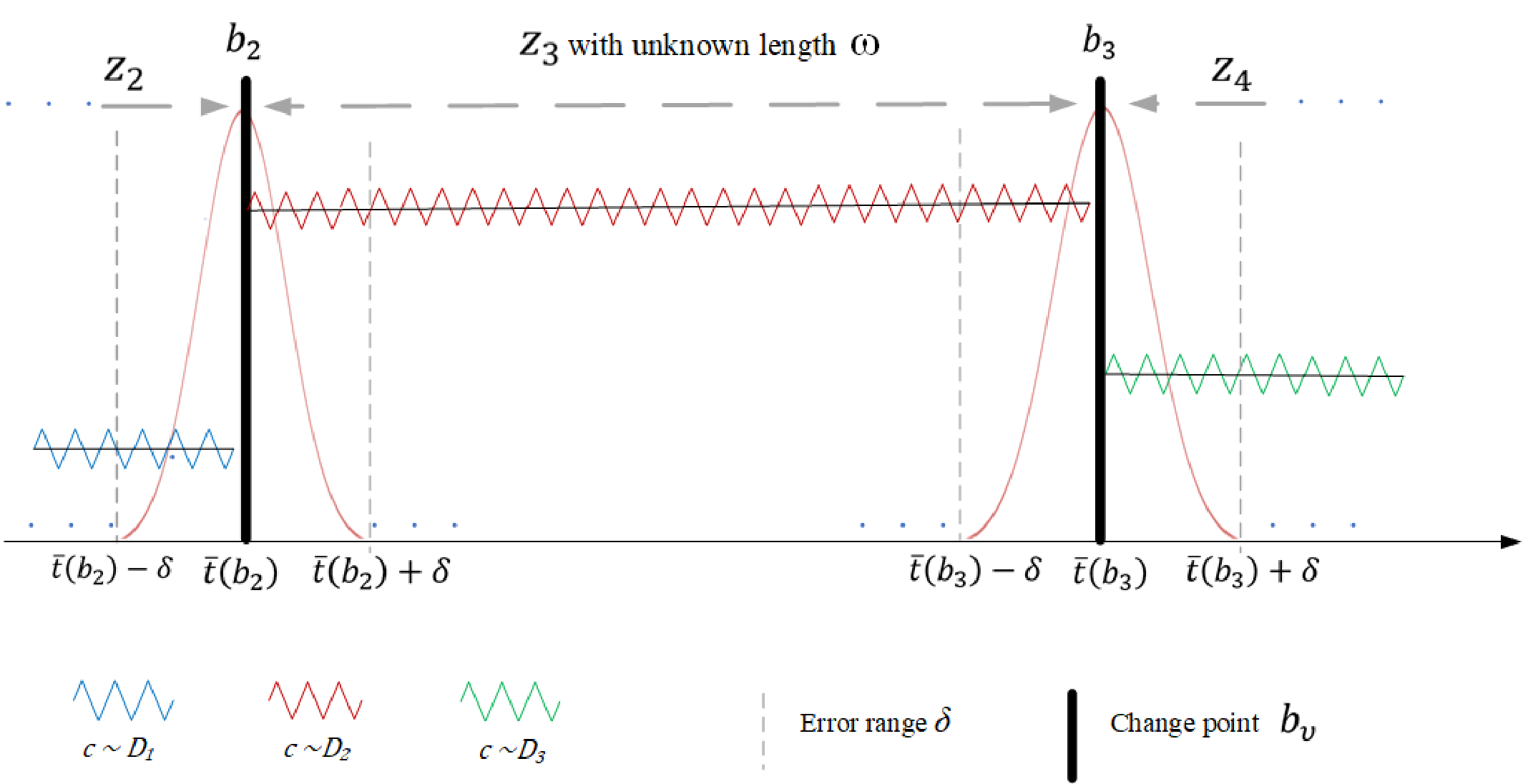}
  \caption{Change point occurrence in a time series.}
 \label{Breakpoint}
 \end{center}
\end{figure}

Let $\mathcal{H}^{\psi}\coloneqq(\nu^{\psi}, \tilde{c}_0, \ldots, \tilde{c}_{\omega}), \ \forall \omega \in z_\psi$ denote the logs or the history on the interval $\psi$ for all $\omega$ rounds. 
We consider that the change points have an unknown distribution with expected inter-arrival of $\mu_{b}$. We make the following assumption.
\begin{assumption}
\label{As:CPD}
{Given $\mathcal{H}^{\psi}$ for all intervals, the observations of the logs reveal $\mu_b$ by an error range of $\delta$}\footnote{This assumption is realistic due to the recent advances in big data analysis that allows extracting and analyzing geometric and statistical patterns of massive size data sets \cite{BigData14}.}.
\end{assumption}
We use the Likelihood Ratio Test (LRT) as an on-line procedure to detect the moment of the change by a sequential change point testing. Based on Assumption \ref{As:CPD}, the logs in the off-policy procedure allow us to obtain the range of the change point occurrence and to narrow down the period in which the LRT is performed. Fig.~\ref{Breakpoint} depicts an example of the cost distribution of one network, which changes after each change point $b_2$ and $b_3$. The range of the change point occurrence where the LRT test is applied can also be seen in Fig.~\ref{Breakpoint}. The horizontal lines on $c$ represent the expected cost of the network in the interval.
 The expected occurrence time of the $\upsilon$-th change point can be written as  $\bar{t}(b_{\upsilon})={\upsilon}\cdot\mu_{b}$. Therefore, the occurrence range of the $\upsilon$-th change point yields $[\bar{t}(b_{\upsilon})-\delta \quad \bar{t}(b_{\upsilon})+\delta]$. Exploiting the logs, there exists a point $\bar{t}(b_{\upsilon})-\delta \in \mathbb{N}$ in the $\psi$-th interval such that $D_1=D_2=\ldots=D_{\bar{t}(b_{\upsilon})-\delta}$, where $D_\omega$ is the cost distribution of the $\omega$-th round belonging the the interval $z_\psi$. However, at some point in the future in the range of $2 \delta$, the change point occurs and the LRT will be performed sequentially in such interval as also depicted in Fig. \ref{Breakpoint}. We are interested to test the null hypothesis defined as
\begin{equation}
    \label{Null_Hypo}
    H_0:D_1=\dots=D_{\varsigma-1}=D_{\varsigma}=\ldots,
\end{equation}
against the alternative hypothesis defined as
\begin{equation}
    \label{Alt_Hypo}
    H_1:\exists \varsigma \in \mathbb{N}: D_1=\dots=D_{\varsigma-1} \neq D_{\varsigma}=D_{\varsigma+1}=\ldots,
\end{equation}
 by focusing on the occurrence of one change point in the alternative hypothesis, where $\varsigma$ is the change point. By assuming a truncated normal distribution in the range of change point occurrence \cite{Seq_Break_Detect}, the likelihood ratio of the two hypotheses up to time  $t^{\prime} \in [\bar{t}(b_{\upsilon})-\delta \quad \bar{t}(b_{\upsilon})+\delta]$, where $t^\prime \geq \varsigma$, for testing $H_0:D \in \boldsymbol{D}_0$ against $H_1:D \in \boldsymbol{D}$ yields
\begin{align}
     \label{LRT}
     &\Lambda(X)=\nonumber \\& \frac{\displaystyle\sup_{D \in \boldsymbol{D}_0} \mathcal{L}(D|X_i)}{\displaystyle\sup_{D \in \boldsymbol{D}} \mathcal{L}(D|X_i)}=\frac{\displaystyle\sup_{\hat{\mu}} \displaystyle \prod_{i=1}^{t^\prime} f(\hat{\mu},\sigma^2|X_i) }{\displaystyle\sup_{\hat{\mu_0}, \hat{\mu_1}} \displaystyle \prod_{i=1} ^{\varsigma} f(\hat{\mu}_0,\sigma^2|X_i) \cdot \displaystyle \prod_{i=\varsigma+1} ^{t^\prime} f(\hat{\mu}_1,\sigma^2|X_i)}
 \end{align}
where $f(\cdot|X_i)$ is the truncated normal probability density function, $\hat{\mu}=\frac{\sum_{i=1}^{t^\prime} x_i}{t^\prime}$, $\hat{\mu}_0=\frac{\sum_{i=1}^{\varsigma}x_i}{\varsigma}$,  $\hat{\mu}_1=\frac{\sum_{i=\varsigma+1}^{t^\prime}x_i}{(t^\prime-\varsigma)}$,  $\boldsymbol{D}$ is the parameter space, and $\boldsymbol{D}_0$ is a specified subset of it. Moreover, $X_i$ are the samples taken from the logs.
\begin{remark}
We apply the LRT to the samples taken from the mostly-selected network; however, in general, the test shall be performed for the samples of each of the networks.
\end{remark}
From $\frac{\partial \ln{\mathcal{L}(\hat{\mu}_0,\hat{\mu}_1|X_i)}}{\partial \sigma^2}=0$, we obtain
\begin{equation}
    \sigma^2=\frac{\displaystyle\sum_{i=1}^{\varsigma} (x_i-\hat{\mu}_0)^2+ \displaystyle\sum_{i=\varsigma+1}^{t^\prime} (x_i-\hat{\mu}_1)^2}{t^\prime}, 
\end{equation}
where the {most likely time of the change point of the variance}\footnote {To avoid misbehavior of the likelihood function, $\varsigma$ shall be at least 5 observations apart from the first and the last values in the series \cite{TimeSeries1987}.} is $\varsigma= \operatorname*{argmin}_{5\leq\varsigma\leq t^\prime-5} \{\sigma(\hat{\mu}_0,\hat{\mu}_1)\}$.
Through some algebraic steps, \eqref{LRT} can be rewritten as 
\begin{equation}
\label{LRTSamples}
   \Lambda(X)= \left(\frac{\displaystyle\sum_{i=1}^{\varsigma} (x_i-\hat{\mu}_0)^2+\displaystyle\sum_{i=\varsigma+1}^{t^\prime} (x_i-\hat{\mu}_1)^2}{\displaystyle\sum_{i=1}^{t^\prime} (x_i-\hat{\mu})^2} \right)^{t^\prime/2}.
\end{equation}
The LRT for testing $H_0$ against $H_1$ has the critical region of the form $\{x:\Lambda(x)\leq k\}$, where $k$ is a real number in range [0,\; 1]. The test will be at significance level $\alpha$ if $k$ satisfies
 
$$
  \sup \{ P(\Lambda(x)\leq k;D \in \boldsymbol{D}_0) \}=\alpha.   
$$
That is, low values of LRT imply that the observed result is less likely to occur under $H_0$ than $H_1$; therefore, the null hypothesis shall be rejected.
\subsubsection{Off-Policy Learning}
\label{Off-Policy_Learning}
The historical data $\mathcal{D}\coloneqq(\mathcal{H}^{\psi},\pi^{\psi}_0)_{\psi=1}^{\Psi}$ is given by executing the logging policy $\pi^{\psi}_0$ on interval $\psi$. Our goal is to develop a target policy $\pi_w^{\psi}$ for each interval exploiting the historical data $\mathcal{D}$. Both logging and target policies map the context to the network with the highest probability distribution. Given $n$ independent and identically-distributed samples, we wish to compute the value of the policy $\pi_w^\psi$ on interval $\psi$ as
\begin{align}
    V^*(\pi_w^\psi)&= \mathbb{E} \left[ \sum_{t=0}^{\omega} c_t^m \middle| \pi_w^\psi, D^{\psi}, \nu^{\psi} \right]= 
    \mathbb{E}_{\pi_w^\psi} \big[ c_t^m\big] \nonumber \\ &= \mathbb{E}_{\nu^{\psi} \sim \chi} \mathbb{E}_{m \sim \pi_w^\psi(\cdot | \nu)} \mathbb{E}_{c \sim D^{\psi}(\cdot | m,\nu)} \big[ c^m_t\big].
\end{align}
To estimate the value of policy $\pi_w^\psi$ on interval $\psi$ with any estimator we have the following
\begin{equation}
\label{Esti_Value}
\hat{V}(\pi_w^\psi| \mathcal{D}) \coloneqq \hat{V}(\pi_w^\psi| \mathcal{H}^\psi, \pi_0^\psi).
\end{equation}
A widely-used off-policy estimator is Inverse Propensity Score (IPS) \cite{HorvitzIPS1952, AgarwalOffPolicy2017} that is used in this paper to evaluate the value of a policy. In a general sense, given logging policy $\pi_0$, target policy $\pi_w$, the logged data, and their distributions, the value of the target policy $\pi_w$ based on IPS estimator is
\begin{align}
    \hat{V}_{\text{IPS}}(\pi_w)&= \frac{1}{T} \sum_{t=1}^{T} \sum_{m} \pi_w(m|\nu_t)\frac{\mathbbm{1}_{\{m_t=m\}}}{\pi_0(m_t|\nu_t)} \tilde{c}_t \nonumber \\&=\frac{1}{T} \sum_{t=1}^{T} \frac{\pi_w(m_t|\nu_t)}{\pi_0(m_t|\nu_t)}\tilde{c}_t,\nonumber
\end{align}
where $\tilde{c}_t$, as defined earlier, is the expected cost of the selected network. 
\begin{definition}
\label{Full_Support}
The logging policy $\pi_0$ is said to have full support for $\pi_w$ when $\pi_0(m|\nu)>0 \quad \forall \nu, m$ for which $\pi_w(m|\nu)>0$. 
\end{definition}
In a stationary setting, to ensure that $\hat{V}_{\text{IPS}}(\pi_w)$ is an unbiased estimate of $V^*(\pi_w)$, $\pi_0$ should have full support for $\pi_w$, i.e. assigning non-zero probabilities to every action in every context \cite{UnbiasedContextual}. According to $\eqref{Esti_Value}$, we can write $\hat{V}_{\text{IPS}}(\pi_w^\psi| \mathcal{D})=\hat{V}_{\text{IPS}}(\pi_w^\psi | \mathcal{H}^\psi, \pi_0^\psi)$. We define  
\begin{equation}
    \hat{V}_{\text{IPS}}(\pi_w^\psi | \mathcal{H}^\psi, \pi_0^\psi)=\sum_{t=1}^{\omega} \frac{\pi_w^\psi(m_t|\nu_t)}{\pi_0^\psi(m_t|\nu_t)}\tilde{c}_t.
\end{equation}
Therefore, the optimal policy of the IPS estimator in each interval can be written as
\begin{equation}
\label{ArgMin_OffPolicy}
    \operatorname*{argmin}_{\pi_w^{\psi}}  \hat{V}_{\text{IPS}}(\pi_w^{\psi} | \mathcal{H}^\psi, \pi_0^\psi).
\end{equation}
To estimate the value of the policy $\pi_w$ over all intervals in our non-stationary environment, we define $\hat{V}_{\text{IPS}}(\pi_w)$ as
\begin{equation}
\hat{V}_{\text{IPS}}(\pi_w)=\frac{1}{\Psi}\sum_{\psi=1}^{\Psi} \hat{V}_{\text{IPS}}(\pi_w^\psi),
\end{equation}
which is the average of the IPS estimator from each interval.
Moreover, we define the regret of the off-policy approach as 
\begin{equation}
\label{Regret_off}
R_T^{\text{Off}}= \sum_{\psi=1}^{\Psi} \hat{V}_{\text{IPS}}(\pi_w^\psi) - \sum_{\psi=1}^{\Psi} V^*(\pi_w^\psi),    
\end{equation}
which is the expected loss of the algorithm compared with the optimal network selection.
The two-step off-policy approach is presented in \textbf{Algorithm \ref{Off-Policy_Approach}}, where given the data set $\mathcal{D}$, the policy for the network selection in each interval is developed based on the on-line SW-UCB approach.
\begin{algorithm}[ht]
\caption{The Off-Policy Approach}\label{Off-Policy_Approach}
\footnotesize
\begin{algorithmic}[1]
\STATE \textbf{Input:} $\mathcal{D}$ 
\STATE \textbf{Output:} $\pi_w^{\psi} \quad \forall z_{\psi} \in \mathcal{Z}$
\STATE $\upsilon=0$, $\psi=1$
      \FOR{ t=1: T}
\STATE \text{Calculate:}
\STATE \text{the estimated change point moment} ($\varsigma$), 
\STATE \text{mean before the estimated change point} ($\hat{\mu}_0$), \STATE \text{mean after the estimated change point} ($\hat{\mu}_1$), 
\STATE \text{and mean of null hypothesis} ($\hat{\mu}$)
\STATE \text{Estimate the LRT using} $\eqref{LRTSamples}$
	\IF {$\Lambda(X)$ $<$ $k$}
	\STATE \text{Extract} $\mathcal{H}^{\psi}$, $\pi^{\psi}_0$ \text{from $\mathcal{D}$}
	\STATE \text{Estimate $\pi_w^{\psi}$ using \eqref{ArgMin_OffPolicy}} 
	\STATE $\upsilon=\upsilon+1$, $\psi=\psi+1$
\ENDIF
      \ENDFOR
\end{algorithmic}
\end{algorithm}
\subsubsection{Overview of the Off-Policy Procedure} 
\label{OffPolicyProcedure}
Concerning the off-policy learning, we use the proposed on-line solution as the logging policy ($\pi_0$) to develop the target policy ($\pi_w$). That is, the \emph{SW-UCB} is the policy based on which the logs are obtained\footnote{In our numerical analysis, we simulate \emph{SW-UCB} policy as $\pi_0$ to log the network selection outcomes as the data set $\mathcal{D}$.}. As discussed in Section \ref{Off-Policy_Learning}, given the logged data, the target policy $\pi_w$ for each interval is the solution of the following optimization problem:
\begin{align}
\mathbf{P2}:& \operatorname*{argmin}_{\pi_w^{\Psi}} \Bigg\{ \pi_w^{\Psi} (m|\nu)\cdot \mathbbm{1}_{\{m=1\}} \cdot \sum_{t=1}^{\omega} \frac{\mathbbm{1}_{\{m_t=1\}} }{\pi_0^{\Psi}(m_t|\nu_t)}\cdot \tilde{c}_t \nonumber \\& +\pi_w^{\Psi}(m|\nu)\cdot \mathbbm{1}_{\{m=2\}} \cdot \sum_{t=1}^{\omega} \frac{\mathbbm{1}_{\{m_t=2\}} }{\pi_0^{\Psi}(m_t|\nu_t)}\cdot \tilde{c}_t \nonumber \\& + \pi_w^{\Psi}(m|\nu)\cdot \mathbbm{1}_{\{m=3\}} \cdot \sum_{t=1}^{\omega} \frac{\mathbbm{1}_{\{m_t=3\}} }{\pi_0^{\Psi}(m_t|\nu_t)}\cdot \tilde{c}_t \Bigg\}
\end{align}
subject to
\begin{align}
\mathbf{C2}:& \pi_w^{\Psi} (m|\nu)\cdot \mathbbm{1}_{\{m=1\}}+\pi_w^{\Psi} (m|\nu)\cdot \mathbbm{1}_{\{m=2\}}\nonumber \\& +\pi_w^{\Psi} (m|\nu)\cdot \mathbbm{1}_{\{m=3\}}=1 \label{c2}\\
\mathbf{C3}:& \pi_w^{\Psi} (m|\nu) > 0 \quad \forall m. \label{c3}
\end{align}
Constraint \eqref{c2} guarantees that the sum of the probabilities equals to one. Constraint \eqref{c3} assigns non-zero probability of selection to each network according to Definition \ref{Full_Support}.  The optimization problem \textbf{P2} can be simply solved using standard solvers such as \emph{CPLEX} with the execution time being less than 0.1 second on a modest hardware. Moreover, after detecting a change point, the optimization problem is solved only once at the beginning of the triggered interval, so that its effect on the overall latency is negligible.
\vspace{0.2cm}
\begin{remark}
The performance of the $\pi_w$ depends on several parameters such as number of arms (networks) and the quality of $\pi_0$. Indeed, large number of arms or weak performance of $\pi_0$ increases error probability in $\pi_w$. 
\end{remark}
%
\section{BS Selection and Relaying Mechanism}
\label{BS_Relay}
\subsection{BS Selection}
\label{BS_Sel}
Once the least congested network type is identified, one of the BSs in the network should be selected for offloading. However, as mentioned before, the BSs of the same network type have on average the same waiting time. Moreover, small waiting time is not the sufficient condition for a BS to be the best among the available ones, mainly due to the effect of some other factors such as the sojourn time. 
To guarantee a successful offloading, i.e. the reception of results by the vehicle, we consider the sojourn distance in the coverage of the BS as a parameter when selecting a BS inside the previously-identified least congested network. 

\begin{figure}
\begin{center}
\includegraphics[width=\columnwidth]{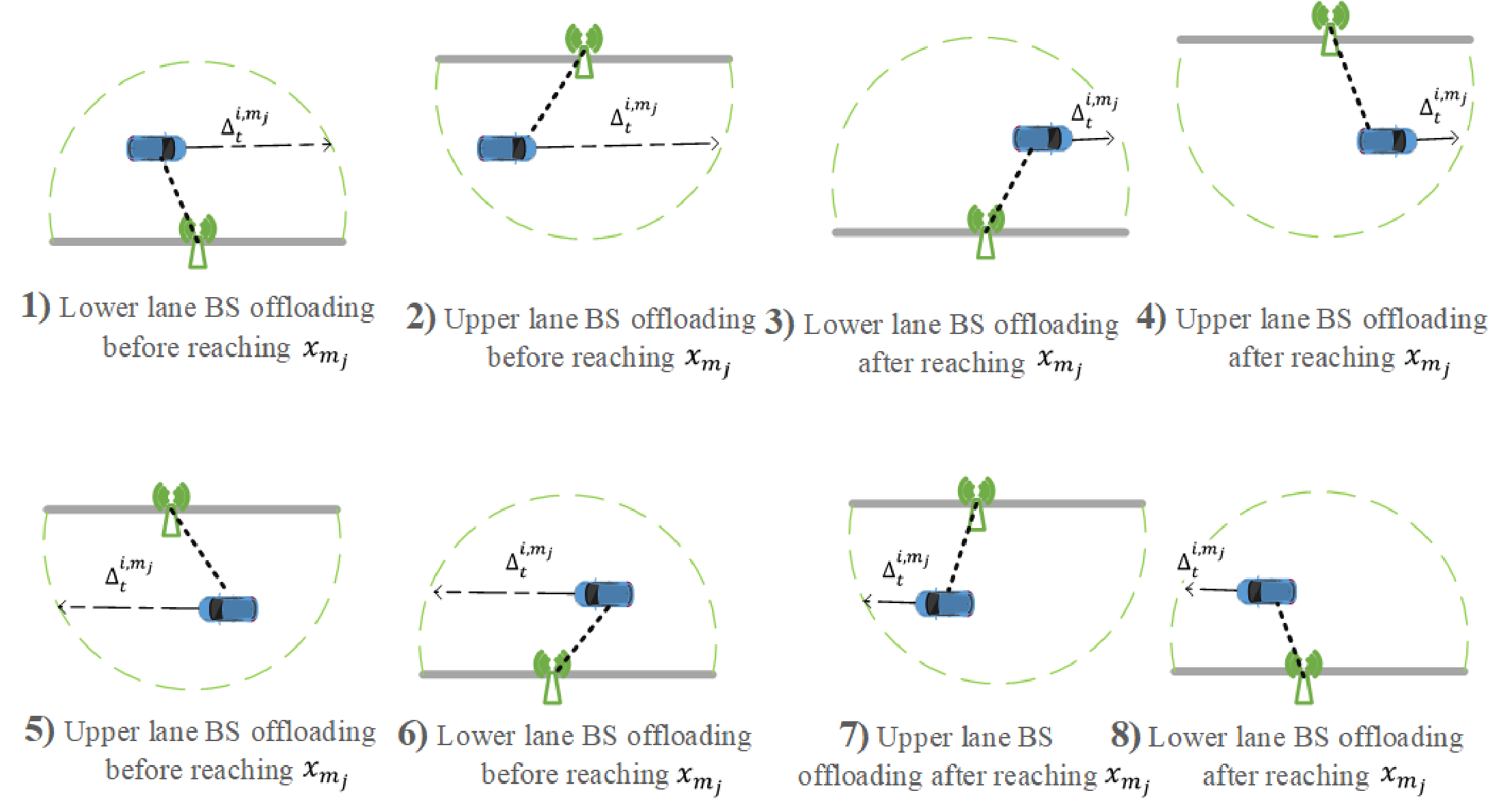}
\caption{Offloading cases considering the locations of devices.}
\label{Mobility_Pattern}
\end{center}
\end{figure}

As shown in Fig.~\ref{Mobility_Pattern}, in a task offloading procedure in our framework, there can be eight offloading cases depending on the locations of the devices. Considering all offloading cases for the $i$th EU, the {sojourn/remaining distance}\footnote{The sojourn distance can be calculated by the EU at any time, since it only requires the knowledge about the locations of the BSs which are fixed.} before going out of the coverage of the $j$th BS at time instant $t$ is equal to~\cite{BozorgchenaniGC2018}
\begin{equation}
	\label{RemDis}
	\Delta_t^{i,{m_j}} =\begin{cases}
   |x_{m_j}-x_{i,t}|+x^\prime & \quad \text{cases 1,2,5,6 in Fig.~\ref{Mobility_Pattern}}  \\
   x^\prime-|x_{m_j}-x_{i,t}| & \quad \text{cases 3,4,7,8 in Fig.~\ref{Mobility_Pattern}},
	\end{cases}
\end{equation}
where
\begin{equation}
    \label{D}
    x^\prime=\sqrt{R_m^2-h_{m_j}^2-|y_{m_j}-y_{i,t}|^2}
\end{equation}
is a distance inside the BS coverage as depicted in Fig.~\ref{Off_Setting}.
As a first step for the BS selection procedure, the EU identifies the BSs in the selected network, as long as it is within their coverage area, as potential candidates for offloading. Hence, we define the set of candidates, $\Im^{F}_t \subset \Im_{j}^m$, in the selected network $m$ that are available for the EU for computation offloading at time instant $t$ as
\begin{equation}
\label{eq:BSList}
\Im^{F}_t= \Big\{N_{m_j}|d_t^{i,m_j} <R_m, \mathbb{Q}^m_t=1\Big\},
\end{equation}
where $d_t^{i,m_j}$is the Euclidean distance between the EU and the $j$th BS of the $m$th network as defined in \eqref{distance}. Moreover, $R_m$ is the coverage area of every BS in network type $m$. Considering the sojourn distance for the BS selection phase, the best BS is selected as
\begin{equation}
\label{BS_selection}
\operatorname*{argmax}_{N_{m_j} \in \Im^{F}_t} \bigg\{ \Delta^{i,m_j}_t \bigg\}. 
\end{equation}
%
\begin{remark}
In case the number of EUs accessing each BS is restricted to some value $\bar{N}$; That is, there is a constraint $\mathbf{C2}: |\Im_{m_j,t}^i| \leq \bar{N}$, where $\Im_{m_j,t}^i=\{ u_i | \mathbb{B}_{m_j}^i=1\}$ and $\mathbb{B}_{m_j}^i$ is an indicator function which returns 1 if EU $i$ offloads to BS $m_j$. Let $\varrho$ be an indicator function that returns 1 if $\mathbf{C2}$ holds and zero otherwise. In this case, the BS is selected as
\begin{equation}
\label{Reward_Unconstrained}
\operatorname*{argmax}_{N_{m_j} \in \Im^{F}_t} \bigg\{ \Delta^{i,m_j}_t \bigg\}\cdot \mathbbm{1}_{ \{\varrho=1\}}. 
\end{equation}
\end{remark}
%
\subsection{Relaying Mechanism}
\label{RelayMechanism}
In VEC, the probability of task loss (or task outage) corresponds to the probability that the EU does not receive an offloaded task in due time, which occurs as a result of the EU mobility. We aim at minimizing the probability of task loss. First, we note that the time that the $i$th EU remains in the coverage area of the $j$th BS (i.e., sojourn time) yields
\begin{equation}
\label{eq:distanceTime}
\widetilde{T}^{i,{m_j}}_t=\frac{\Delta^{i,{m_j}}_t}{v_{i}}.
\end{equation}
Moreover, the overall offloading time is given by
\begin{equation}
\label{T_off}
T_{l,t}^{\text{off},{m_j}}= \frac{ \rho^{\text{up}}_l}{r_{m_j,t}^{\text{up}}}+\frac{O\cdot \rho^{\text{up}}_l}{\eta_{m_j}}+T_{w_m}^{l}(\vartheta_t^m)+\frac{\rho^{\text{dl}}_l}{r_{m_j,t}^{\text{dl}}}.
\end{equation}
For the $l$th task generated at time instant $t$, the task loss can be formalized as
\begin{equation}
	\label{eq:pktloss}
	\Omega_{l,t}^{m_j} =\begin{cases}
   1 & \quad \text{if } \widetilde{T}_t^{i,{m_j}}<T_{l,t}^{\text{off},{m_j}}\\
   0 & \quad \text{if } \widetilde{T}_t^{i,{m_j}}\geq T_{l,t}^{\text{off},{m_j}}
	\end{cases}.
\end{equation}
In words, a task loss occurs if the total offloading latency is larger than the EU sojourn time within the coverage area of the serving BS. The EU aims at minimizing the task loss during the time horizon $T$, i.e., $\minimize \sum_{t=1}^T \Omega_{l,t}^{m_j}$.
Even for the best BS w.r.t. the sojourn distance, the overall offloading latency might be more than the sojourn time.
To address this challenge, we develop a relaying mechanism.

Let the \textit{original BS} refer to the BS selected for task offloading. The \textit{neighboring BS} is a BS towards which (i.e. its coverage area) the vehicle moves. A relaying mechanism allows a vehicle to collect the task result from the neighboring BS, if receiving the result from the original BS is not possible. That is, if the BSs can communicate with each other, the task is relayed through the backhaul network \cite{MECVehicular17}. Then the relayed offloading time yields
\begin{equation}
\label{T_off_Relay}
\hat{T}_{l,t}^{\text{off},{m_j}}= \frac{ \rho^{\text{up}}_l}{r_{{m_j,t}}^{\text{up}}}+\frac{O\cdot \rho^{\text{up}}_l}{\eta_{m_j}}+T_{w_m}^l(\vartheta_t^m)+ T_l^{\text{trans},Op} +\frac{\rho^{\text{dl}}_l}{r_{{m_k,t}}^{\text{dl}}},
\end{equation}
where $T_l^{\text{trans,Op}}$ is the transmission time for relaying through backhaul, and $\frac{\rho^{\text{dl}}_l}{r_{m_k,t}^{\text{dl}}}$ the reception time from the destination BS. We then redefine the outage as
\begin{equation}
	\label{pktloss_Relay}
	\hat{\Omega}_{l,t}^{m_j} =\begin{cases}
   1 & \quad \text{if } \widetilde{T}^{i,{m_j}}_t<\frac{ \rho^{\text{up}}_l}{r_{{m_j,t}}^{\text{up}}}\\
   0 & \quad \text{if } \widetilde{T}^{i,{m_j}}_t\geq \frac{ \rho^{\text{up}}_l}{r_{{m_j,t}}^{\text{up}}}
	\end{cases}.
\end{equation}
This implies that when employing a relaying mechanism, a task outage occurs only if the transmission time to the original BS is not sufficient.

\section{Performance Evaluation}
\label{sec:numericalresult}
%
We evaluate the performances of the proposed learning methods by simulations.
\subsection{Time-Scale of Operations of the Learning Algorithms}
\label{Alg_Sec}
In order to have a better understanding of the scenario considered for simulation results, we elaborate here the time-scale of the two proposed learning algorithms, each composed of a decision-making and an offloading phase.
\begin{figure}[tb]%
\centering
 \subfloat[The time-scale of the two phases for the on-line approach.]{\includegraphics[width=\columnwidth]{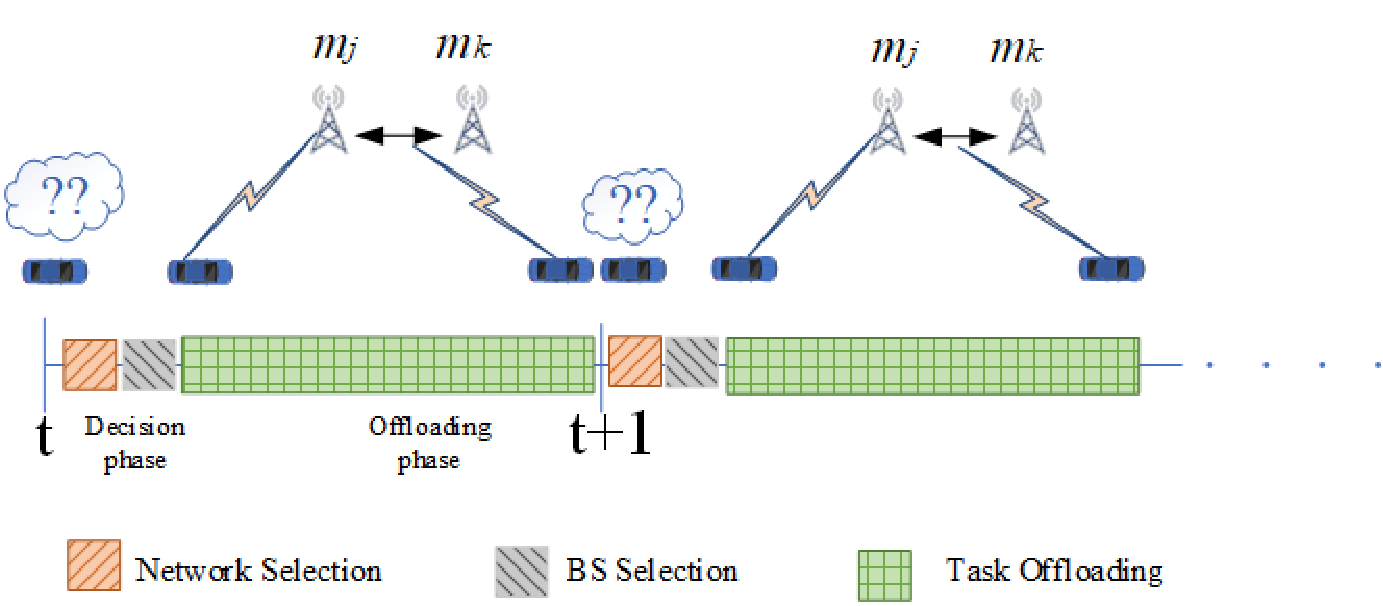}\label{onFigure}}\\
\subfloat[The time-scale of the two phases for the off-policy approach.]{\includegraphics[width=\columnwidth]{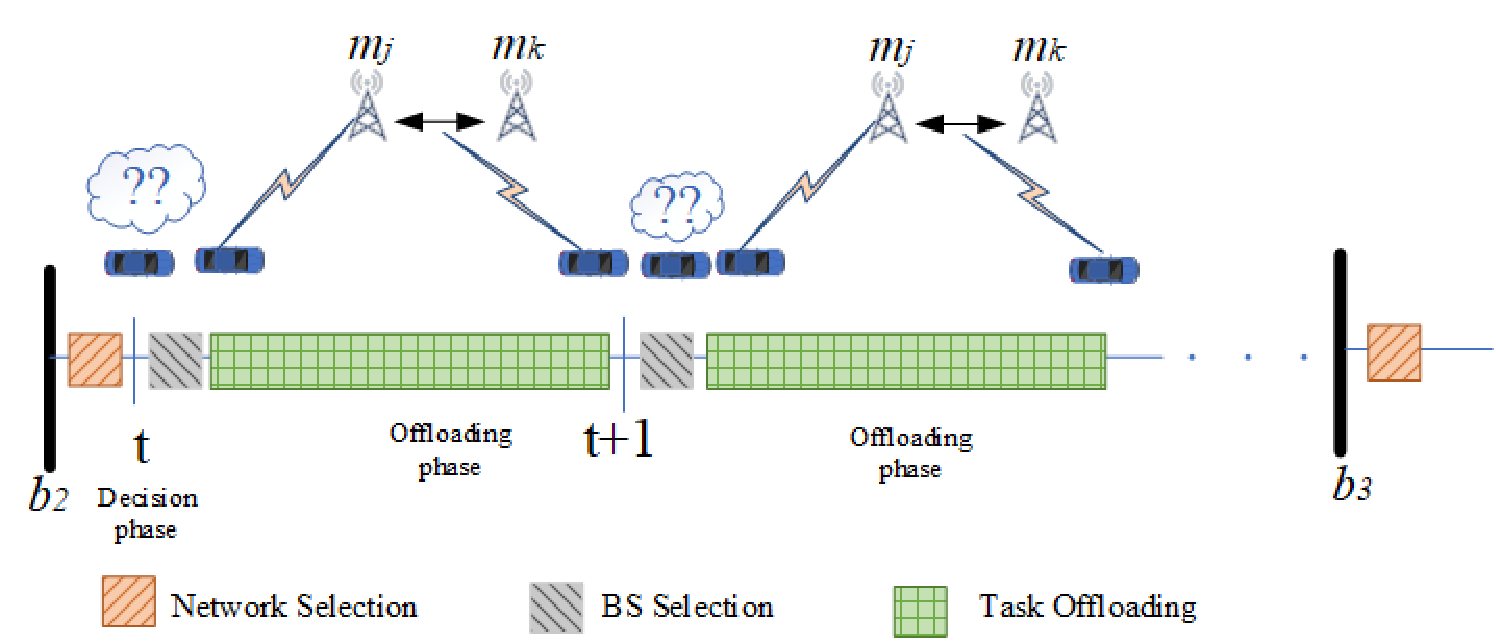}\label{offFigure}}%
\caption{The time-scale of the two phases for the two learning approaches.}
\label{AlgIllus}
\end{figure}

As illustrated in Fig.~\ref{AlgIllus}, in the first phase, the EU decides where to offload. This decision phase includes network selection, as explained in Section~\ref{Net_Selection}, and BS selection, as explained in Section~\ref{BS_Sel}. Afterward, the user offloads the task to the selected BS and later receives the result either from the \textit{original} BS or the \textit{neighboring} BS, in case of relaying, as explained in Section~\ref{RelayMechanism}.

The on-line approach includes the network selection and BS selection at every round of decision-making, i.e., upon generation of a task, as illustrated in Fig.~\ref{onFigure}. After selecting the BS, the offloading starts. The off-policy approach, in contrast, performs the network selection only upon detecting a change point, as explained in Section~\ref{Off-Policy_Learning}\footnote{Change point detection algorithm is run within the range of $2\delta$ as explained in Section~\ref{ChangepointSection}.}. However, upon generation of a task at every round, only the BS selection is performed. Similar to the on-line approach,  the offloading phase starts after selecting the BS. It should be noted that the learning solutions do not incur any signaling overhead, since there is no message exchange between the vehicles and the BSs in the learning procedure.

\subsection{Simulation Parameters}
To evaluate the proposed offloading schemes, we perform numerical analysis. Table~\ref{tab1} summarizes the simulation parameters. The vehicle has a random location in the area. There are three networks (e.g. Macro, Micro, and Pico cellular networks) available in the area. The Macro and Micro BSs are placed in the upper and lower side of the road and the Pico BSs in the upper, lower and middle of the road all following uniform distribution. By this placement of BSs, at least one network is available to serve the vehicle  at all time. There are 25000 simulation rounds out of which we consider only the rounds in which all the three network types are available to guarantee a fair comparison.

Table~\ref{tab2} presents the traffic of each network in different intervals. A vehicle generates a task sequentially. We consider two applications: (i) A processing application generating tasks that requires a higher number of processing operations (e.g. image processing); (ii) A collecting application that requires a lower number of processing operations, (e.g. sensor data analysis). In Table~\ref{Task_Par}, we summarize the numerical values, expressed in terms of Floating Point Operations (FLOP) per task data size~\cite{BozorgTMC20}.
\begin{table}[tbp]
\centering
\caption{Simulation Parameters}
\label{tab1}
\scriptsize
\begin{tabular}{ | m{4.0cm} | m{3.5cm}| } 
  \hline
  \textbf{Parameter} & \textbf{Value} \\
  \hline
	\hline
	Height of Ma-BS  & 20 $m$ \cite{ITU}\\
	\hline
     Height of Mi-BS & 6 $m$ \cite{ITU}\\
    \hline
     Height of Pi-BS &  3 $m$ \cite{ITU}\\
	\hline
	Path-loss attenuation factor ($\beta_t^{i,{m_j}}$) &  $140.7+36.7 \log_{10}\left(\frac{d_t^{i,{m_j}}}{1000}\right)$\cite{3GPP}\\
  \hline
   Bandwidth ($B_{m}$)& \SI{10}{\mega\hertz} \cite{LearningOffEdge}\\
	\hline
	Noise power density ($\zeta$)& $10^{-13} B_{m}$ \cite{LearningOffEdge}\\
	\hline
	Ma-BS coverage range & 500 $m$ \cite{ITU}\\
	\hline
	Mi-BS coverage range & 200 $m$ \cite{ITU}\\
	\hline
	Pi-BS coverage range  & 100 $m$ \cite{ITU}\\
	\hline
	EU computational power ($\eta_{i}$)& 15 G FLOPS\\
	\hline
	Ma computational power & $4*\eta_{i}$\\
	\hline
	Mi computational power & $3*\eta_{i}$\\
	\hline
	Pi computational power & $2*\eta_{i}$\\
	\hline
	Significance level ($\alpha$) & 0.05\\
	\hline
	EU velocity ($v_i$)& 10-20 meters/second \\
	\hline
	Error range in LRT ($\delta$) & 500 rounds\\
	\hline
	Inter-arrival of change points ($\mu_b$) & 5000 rounds\\
	\hline
\end{tabular}
\end{table}
\begin{table}[tbp]
\centering
\caption{MAB Setting}
\label{tab2}
\scriptsize
\begin{tabular}{ | m{1.7cm} | m{1.8cm}|  m{1.8cm}| m{1.8cm}| } 
  \hline
  \textbf{ } & \textbf{Macro Network} & \textbf{Micro Network} & \textbf{Pico Network}\\
  \hline
	\hline
  Intervals & \pbox{45cm}{t=1$\sim$ 5000\\ t=5000$\sim$ 10000\\ t=10000$\sim$ 15000\\ t=15000$\sim$ 20000\\ t=20000$\sim$ 25000\\ } & \pbox{45cm}{t=1$\sim$ 5000\\ t=5000$\sim$ 10000\\ t=10000$\sim$ 15000\\ t=15000$\sim$ 20000\\ t=20000$\sim$ 25000\\ } & \pbox{45cm}{t=1$\sim$ 5000\\ t=5000$\sim$ 10000\\ t=10000$\sim$ 15000\\ t=15000$\sim$ 20000\\ t=20000$\sim$ 25000\\ }\\
	\hline
	 Traffic model parameters & \pbox{40cm}{$\lambda$=7 , $\mu$=0.4 \\ $\lambda$=3 , $\mu$=0.3 \\ $\lambda$=10 , $\mu$=0.5 \\ $\lambda$=3 , $\mu$=0.3 \\ $\lambda$=7 , $\mu$=0.4} & \pbox{40cm}{$\lambda$=3 , $\mu$=0.3 \\ $\lambda$=10 , $\mu$=0.5 \\ $\lambda$=7 , $\mu$=0.4 \\ $\lambda$=10 , $\mu$=0.5 \\ $\lambda$=10 , $\mu$=0.5} & \pbox{40cm}{$\lambda$=10 , $\mu$=0.5 \\ $\lambda$=7 , $\mu$=0.4 \\ $\lambda$=3 , $\mu$=0.3 \\ $\lambda$=7 , $\mu$=0.4 \\ $\lambda$=3 , $\mu$=0.3}\\
	\hline
	Expected cost & \pbox{60cm}{$c_t^m=2.8$ \\ $c_t^m=0.9$ \\ $c_t^m=5$ \\ $c_t^m=0.9$ \\ $c_t^m=2.8$} & \pbox{60cm}{$c_t^m=0.9$ \\ $c_t^m=5$ \\ $c_t^m=2.8$ \\ $c_t^m=5$ \\ $c_t^m=5$} & \pbox{60cm}{$c_t^m=5$ \\ $c_t^m=2.8$ \\ $c_t^m=0.9$ \\ $c_t^m=2.8$ \\ $c_t^m=0.9$}\\
	\hline
\end{tabular}
\end{table}
\begin{table}[tbp]
\centering
\caption{Task Parameters}
\label{Task_Par}
\begin{tabular}{ | m{0.53\columnwidth} | m{0.35\columnwidth}|} 
  \hline
  \textbf{Task Parameter} & \textbf{Value}\\
  \hline 
  	Task size ($\rho^{\text{up}}_l$) & [1 5] MB\\
	\hline
	Offloaded to downloaded portion  & 5\\
	\hline
  Processing application operations & 10 G FLOP per MB\\
  \hline
  Collecting application operations & 1 G FLOP per MB\\
  \hline
\end{tabular}
\end{table}
\subsection{Impact of UCB Parameters on the On-line Approach}
We first investigate the impact of parameters on the performance of \emph{SW-UCB}. 
\begin{figure}[tb]
\centering
\color{black} \subfloat[Impact of $\tau$ on average regret.]{\includegraphics[width=0.9\columnwidth]{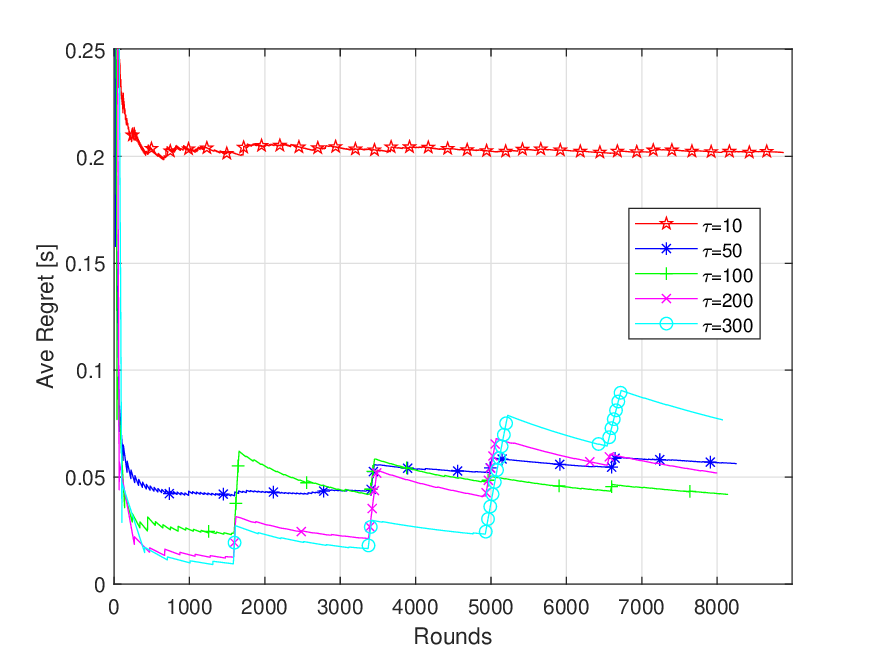}\label{RegretTW}}\\
\subfloat[Impact of $\tau$ on sub-optimal network selection.]{\includegraphics[width=0.8\columnwidth]{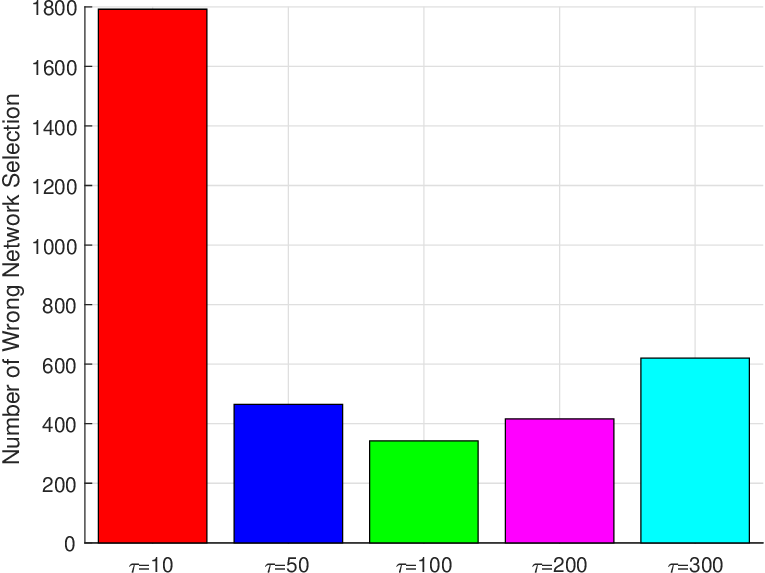}\label{ArmTW}}%
\caption{Impact of the window length $\tau$. }
\label{TW_Figure}
\end{figure}
Fig.~\ref{TW_Figure} shows the impact of window length on both number of sub-optimal network selection and the average regret. Sub-optimal network selection is mainly affected by the window length. With a small window length the change point is detected faster, whereas a large window size might result in detection delay. Nonetheless, if the window is too small, there might not exist sufficient historical data for optimal decision-making during the interval between two change points. Based on the experiments, we select the window length as $\tau=100$ rounds for rest of the simulation as it results in the best network selection, thereby the lowest average regret. %
\begin{figure}[ht]
\color{black}
\begin{center}
\includegraphics[width=0.95\columnwidth]{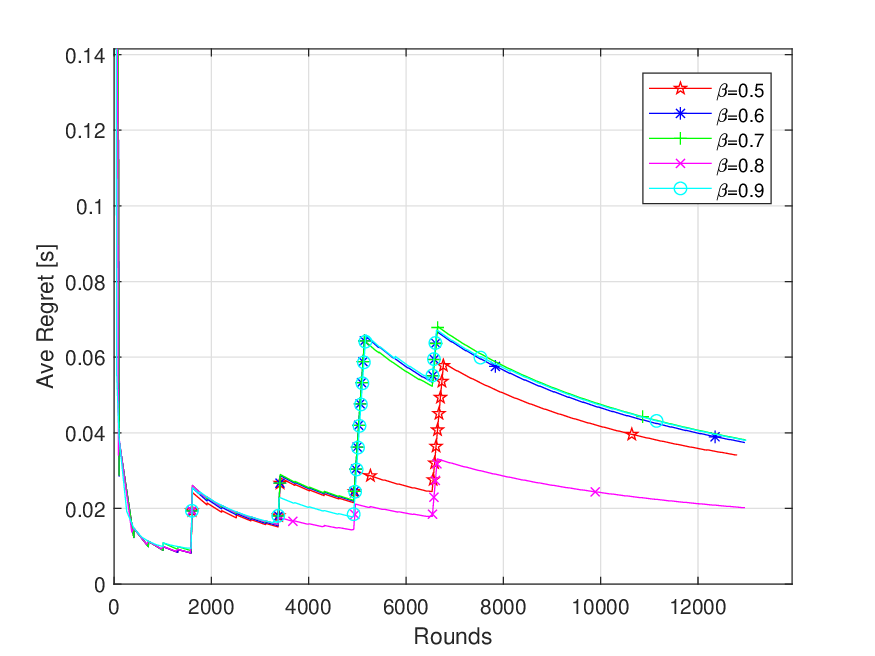}
\caption{Impact of $\beta$ on average regret.}
\label{RegretBeta}
\end{center}
\end{figure}
\begin{figure}[ht]
\color{black}
\begin{center}
\includegraphics[width=\columnwidth]{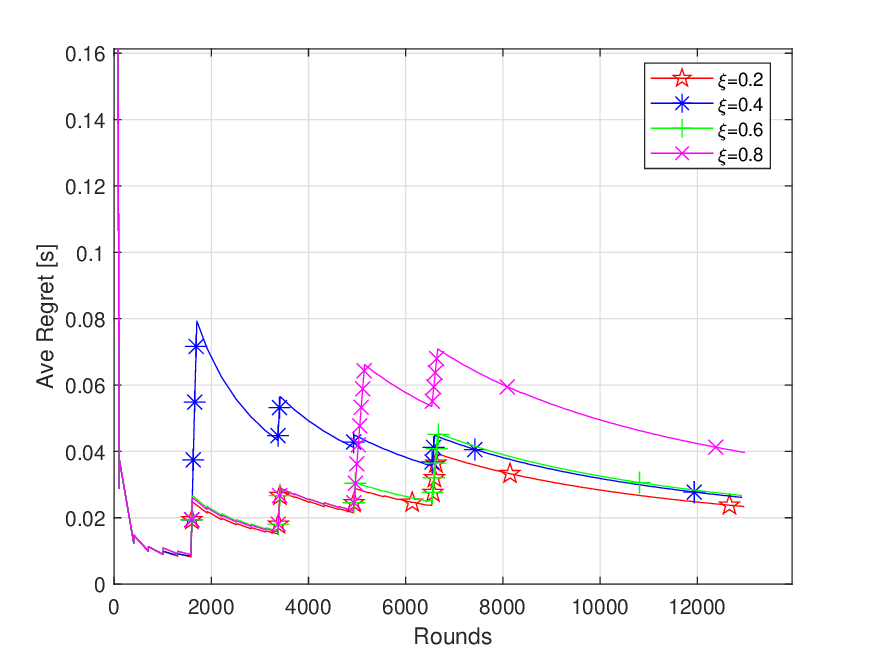}
\caption{Impact of $\xi$ on average regret.}
\label{RegretXi}
\end{center}
\end{figure}
\begin{figure}[ht]
\color{black}
\begin{center}
\includegraphics[width=1.1\columnwidth]{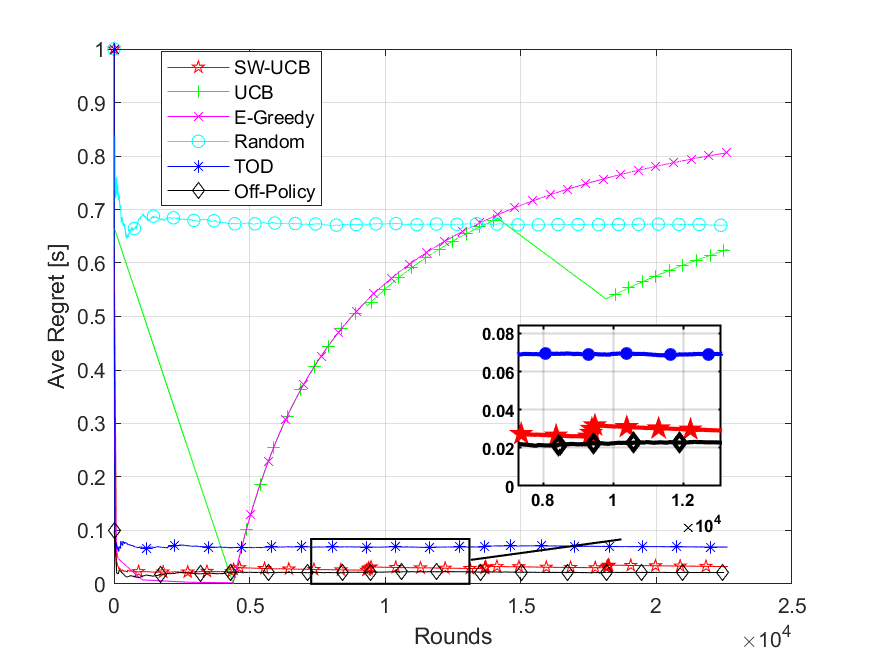}
\caption{Average regret.}
\label{AveRegret}
\end{center}
\end{figure}
\begin{figure*}[ht]
\begin{center}
\includegraphics[width=\textwidth]{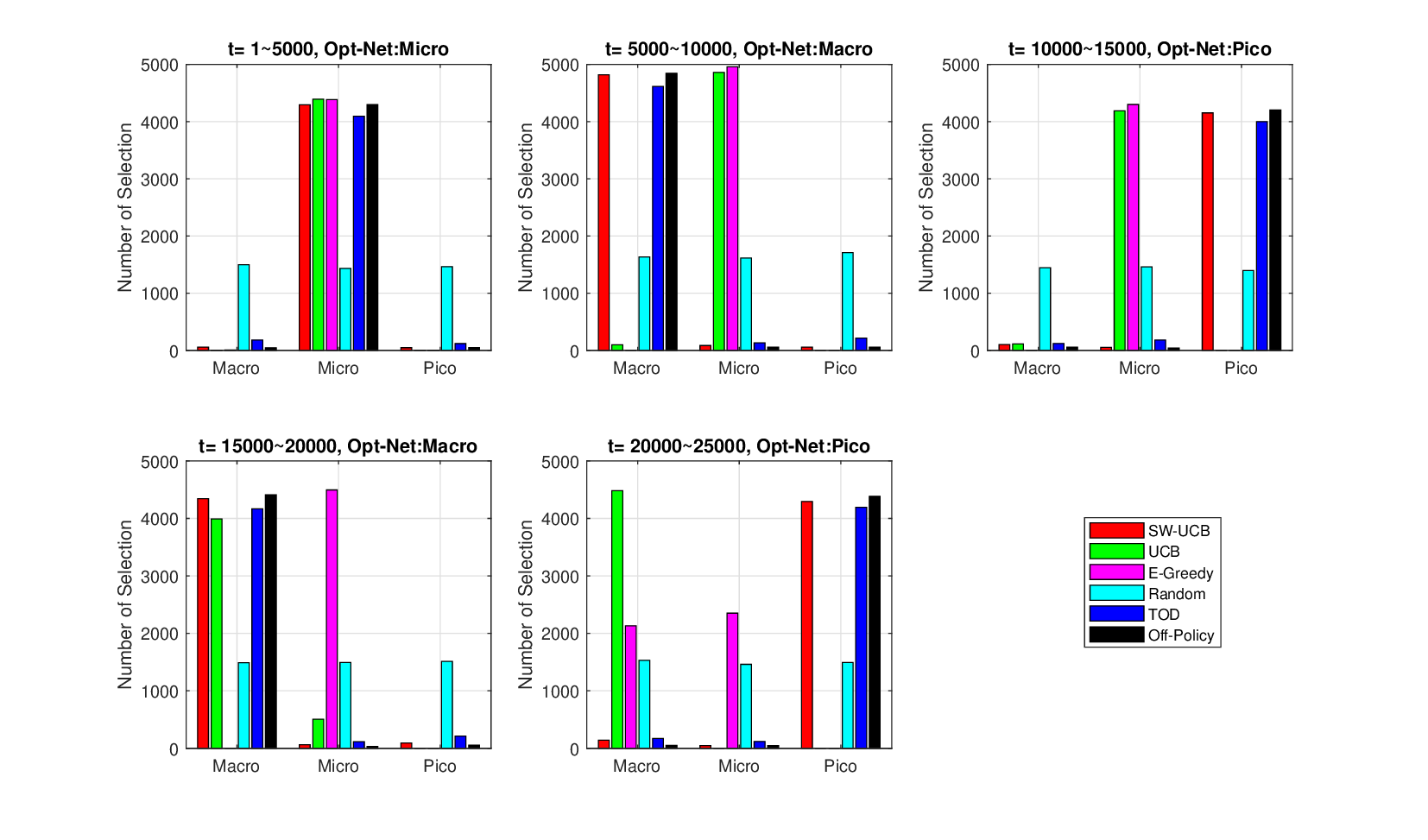}
\caption{Network selection in each interval.}
\label{ArmSelection}
\end{center}
\end{figure*}

We also study the impact of $\beta$, i.e. the coefficient of the exploration factor in the UCB index, on the average regret. Fig.~\ref{RegretBeta} depicts the results. We select $\beta=0.8$ for the rest of the simulation due to its superior performance. Furthermore, the effect of $\xi$ on the average regret is shown in Fig.~\ref{RegretXi}. From the figure, the value of $\xi=0.2$ has the best performance.
\subsection{Comparison With Other Approaches}
We compare the performance of \emph{SW-UCB} and \emph{Off-policy} solutions with the following benchmarks:
\begin{itemize}
 \item \emph{UCB}: 
 The seminal upper-confidence bound policy~\cite{UCBAgarwal1995};
\item \emph{$\epsilon$-greedy}: At each round $t$, it selects a random arm with probability $\epsilon=1/t$ and the best arm so far with probability 1-$\epsilon$.
\item \emph{Random}: At each round, it selects an arm uniformly at random.
 \item \emph{TOD}: It selects the network based on an approach belonging to the UCB family, namely, \textit{discounted-UCB}. This approach assigns an index to each choice at every round of decision-making. In calculating the index, recent observations play a significant role as they receive high weights, and the effect of observations diminishes over time \cite{LearnPick18}.
\end{itemize}
Fig.~\ref{AveRegret} illustrates the performance of different approaches in terms of average regret. The proposed \emph{SW-UCB} approach shows the best performance compared to other benchmarks. \emph{TOD} approach is the closest to our proposed solutions. \emph{$\epsilon$-greedy} and \emph{UCB} exhibit similar performance which is inferior in comparison with other methods. \emph{Random} selection leads to a severe sub-optimal network selection. \emph{Off-policy} approach has the lowest regret due to change point detection mechanism and the proposed network selection method.

Fig.~\ref{ArmSelection} represents the network selection of the agent in different intervals while using different approaches. The parameters regarding the interval length and the expected cost in each interval are gathered in Table~\ref{tab2}. In the first interval, four approaches, namely \emph{SW-UCB}, \emph{UCB}, \emph{$\epsilon$-greedy}, and \emph{Off-Policy} show similar performance as they continue pulling the optimal arm. Nonetheless, after the change point, i.e. when the optimal network changes to Macro, their performances become different. The \emph{Off-Policy} approach shows the best performance due to the LRT for change point detection and the developed $\pi_w$. Moreover, \emph{SW-UCB} is able to detect the change and thus performs well. \emph{TOD} has the closest performance to \emph{SW-UCB} due to using the discounted-UCB algorithm. The \emph{UCB} method considers the entire history for decision-making; therefore, it requires a long time to adapt to the change. Similarly, \emph{$\epsilon$-greedy} only exploits the historical knowledge and thus continues selecting the Micro network (exploration loses its color by time). \emph{Random} approach does not take the changes in cost distribution into account and hence has the worst performance. All approaches show the same behavior in the other intervals. 
\begin{figure*}[h]
\begin{center}
\includegraphics[width=\textwidth]{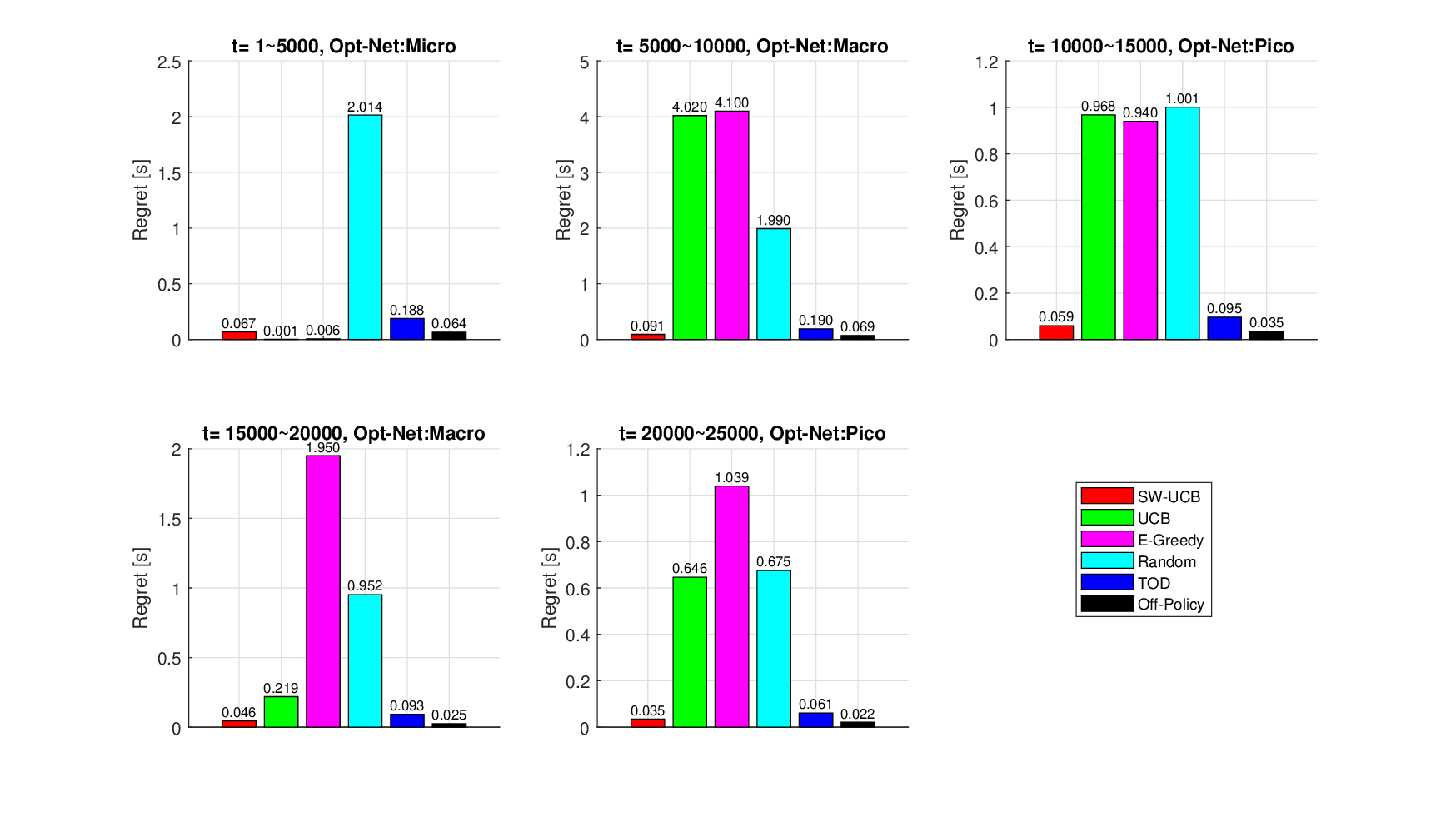}
\caption{Average regret in each interval.}
\label{RegretInterval}
\end{center}
\end{figure*}

Fig.~\ref{RegretInterval} depicts the average regret, which is determined by the frequency of sub-optimal network selection and also the selected network. The average regret of \emph{SW-UCB} and \emph{Off-Policy} approaches are the smallest compared to the other methods.
\begin{figure}[h]
\begin{center}
\includegraphics[width=0.9\columnwidth]{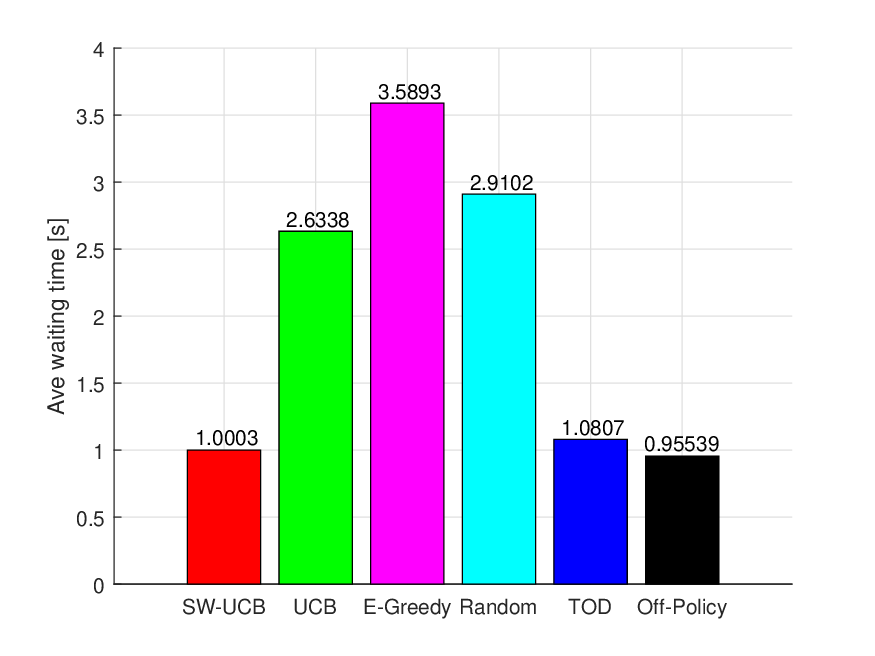}
\caption{Average waiting time for an offloaded task.}
\label{T_W}
\end{center}
\end{figure}
Fig.~\ref{T_W} depicts the average waiting time each task experiences at the selected BS. Clearly, by selecting the network with the lowest congestion in each interval, the tasks suffer lower waiting time. The proposed \emph{SW-UCB} and \emph{Off-Policy} approaches have the lowest waiting time for the tasks, which is the result of the optimal network selection in different congestion patterns.
\begin{figure}[h]
\begin{center}
\includegraphics[width=0.95\columnwidth]{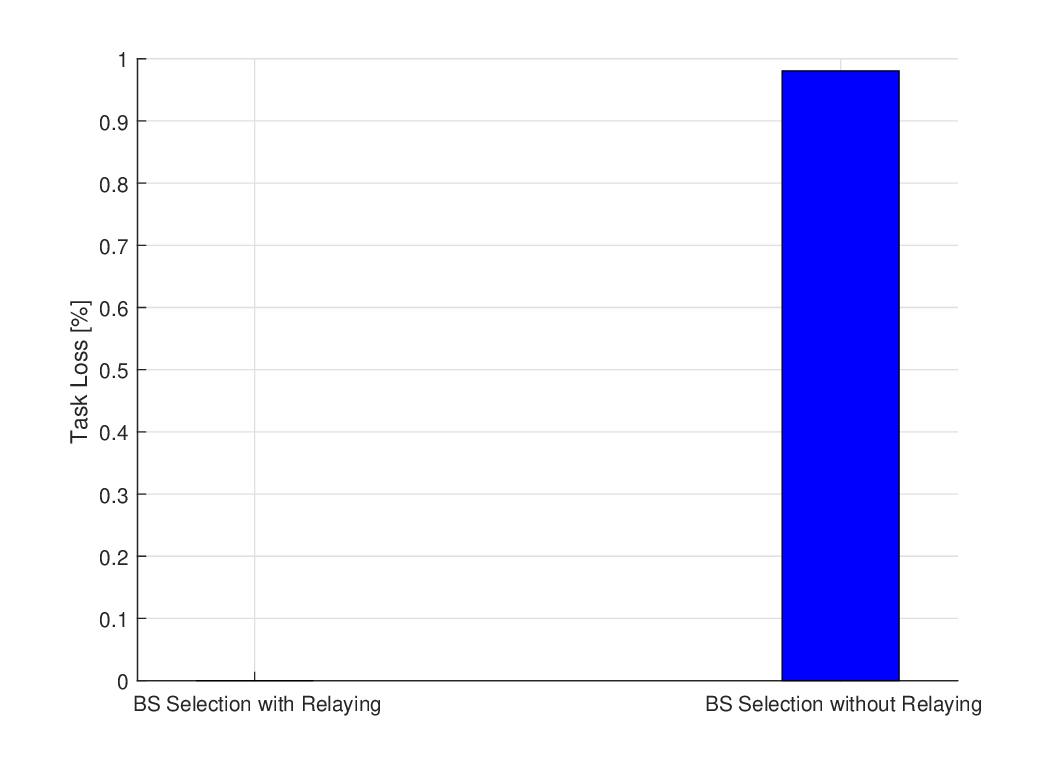}
\caption{Task loss with and without relaying mechanism.}
\label{Pkt_Loss}
\end{center}
\end{figure}
\textcolor{black}{Finally, Fig.~\ref{Pkt_Loss} shows that the relaying mechanism improves the performance in terms of the task loss and brings it to a value close to zero.} Indeed, the residual loss in the absence of a relaying mechanism is mainly due to the small sojourn time that is insufficient even for transmission of the task to the original BS. The relaying mechanism mitigates this  task loss.
\section{Conclusion}
\label{conclusion}
We have studied a task offloading problem in a VEC scenario where several wireless access networks with dynamic traffic patterns co-exist. We have proposed a multi-arm bandit approach, namely, the \emph{SW-UCB} approach, to solve the network selection problem. We have also proposed an off-policy approach to detect the change points and select the best network. Moreover, we have developed a BS selection and a relaying mechanism that reduces the waiting time and task loss. Numerical results have demonstrated the effectiveness of the proposed approaches in selecting the least congested network and adapting to the changes in the network traffic. The proposed \emph{SW-UCB} has a lower average regret and lower task latency than other benchmarks. Moreover, the \emph{Off-Policy} approach has the lowest regret over rounds and average task waiting time than all the other approaches.

As a future work, we plan to extend the single-agent MAB scenario to a multi-agent setting. In such a scenario, the traffic in each network depends on the joint decisions of agents that complicates the problem. Moreover, we would like to consider the problem of cost optimization of the system in terms of number/density of activated BSs required to maintain the latency for task offloading to an acceptable level. \textcolor{black}{Furthermore, we plan to study the impact of energy consumption on the vehicular environment and consider a joint latency and energy consumption optimization.}
\bibliographystyle{IEEEtran}
\bibliography{IEEEabrv,Biblio}

\end{document}